\documentclass[12pt]{article}
\usepackage{graphicx}
\usepackage[T1]{fontenc}
\usepackage{color}
\usepackage{bm}
\usepackage{tabularx}
\usepackage{tikz}
\usepackage{mathtools}
\usepackage{floatrow}
\usepackage{multirow}
\usepackage{caption}
\usepackage{natbib}
\RequirePackage{amsthm,amsmath,amsfonts,amssymb}
\RequirePackage[colorlinks,citecolor=blue,urlcolor=blue]{hyperref}
\RequirePackage{graphicx}
\usepackage{subcaption}
\usepackage{lipsum}
\usepackage{commath}

\newcolumntype{R}{>{\raggedleft\arraybackslash}X}

\newcommand{\iid}{\text{i.i.d}}

\newcommand{\Poi}{\text{Poi}}

\newcommand\floor[1]{\lfloor#1\rfloor}

\newcommand{\R}{\mathbb{R}}

\def\P{\mathbb{P}}
\def\E{\mathbb{E}}

\newcommand{\N}{\mathbb{N}}

\newcommand{\X}{\mathcal{X}}

\newcommand{\IL}{I_{\mathcal{L}}}
\newcommand{\ind}[1]{\mathbf{1}\big\{#1\big\}}

\newcommand{\im}{\mathrm{im}}

\DeclarePairedDelimiterX{\inp}[2]{\langle}{\rangle}{#1, #2}

\definecolor{black}{gray}{0}
\definecolor{gray1}{gray}{0.25}
\definecolor{gray2}{gray}{0.5}
\definecolor{gray3}{gray}{0.75}
\definecolor{white}{gray}{1}

\theoremstyle{plain}
\newtheorem{theorem}{Theorem}[section]

\newtheorem{proposition}[theorem]{Proposition}

\theoremstyle{definition}

\newtheorem{remark}[theorem]{Remark}

\begin{document}

\def\spacingset#1{\renewcommand{\baselinestretch}%
{#1}\small\normalsize} \spacingset{1}

\title{\bf Feature Detection and Hypothesis Testing for Extremely Noisy Nanoparticle Images using Topological Data Analysis}
  \author{
    Andrew M. Thomas\thanks{
    The authors would like to acknowledge Joshua Vincent, Piyush Haluai and Mai Tan at Arizona State University (ASU) for sample preparation and TEM data acquisition. We also thank Ramon Manzorro (University of C{\'a}diz) and Advait Gilankar (ASU) for assistance in imaging simulations and data handling as well as the John M. Cowley Center for High Resolution Electron Microscopy at ASU. The authors gratefully acknowledge funding from NSF grants OAC-1940124, CCF-1934985, DMS-2114143 and specifically CBET-1604971 and OAC-1940263 for the support of sample preparation and data acquisition.} \hspace{10cm}\\
    Center for Applied Mathematics, Cornell University, \\
    Peter A. Crozier \\
    School for Engineering of Matter, Transport and Energy, ASU, \\
    Yuchen Xu \\ 
    Department of Statistics and Data Science, Cornell University, \\
    David S. Matteson \\
    Department of Statistics and Data Science, Cornell University}
\date{}
\maketitle







\begin{abstract}
We propose a flexible algorithm for feature detection and hypothesis testing in images with ultra low signal-to-noise ratio using cubical persistent homology. Our main application is in the identification of atomic columns and other features in transmission electron microscopy (TEM). Cubical persistent homology is used to identify local minima and their size in subregions in the frames of nanoparticle videos, which are hypothesized to correspond to relevant atomic features. We compare the performance of our algorithm to other employed methods for the detection of columns and their intensity. Additionally, Monte Carlo goodness-of-fit testing using real-valued summaries of persistence diagrams derived from smoothed images (generated from pixels residing in the vacuum region of an image) is developed and employed to identify whether or not the proposed atomic features generated by our algorithm are due to noise. Using these summaries derived from the generated persistence diagrams, one can produce univariate time series for the nanoparticle videos, thus providing a means for assessing fluxional behavior. A guarantee on the false discovery rate for multiple Monte Carlo testing of identical hypotheses is also established.  
\end{abstract}

\noindent%

\section{Introduction}

Transmission electron microscopy (TEM) has become a critical tool in both physical and life sciences for characterizing materials at the atomic level. Over the last 10 years, recent advances in direct electron detectors have greatly improved sensitivity with detective quantum efficiencies approaching the theoretical maximum of unity \citep{ruskin2013, faruqi2018, levin2021direct}. As a result, the information content in signals is now limited mostly by the shot noise (Poisson noise) associated with the quantum mechanical processes responsible for electron emission and scattering. In principle, for a perfect detector, the fraction of noise in the signal can be made arbitrarily small by counting for longer or by increasing the flux of the incident electron beam. However, increasing the measurement time/electron flux is simply not practical for many materials systems, as they are irreversibly damaged by the electron beam \citep{egerton2004, egerton2013, egerton2019}. 

The signal-to-noise ratio is also significantly limited when high temporal resolution is required for investigations of dynamic behavior associated with kinetic processes in materials. In such experiments, the exposure time per frame is necessarily short resulting in a high degree of shot noise in each frame. For example, recent efforts to understand structural dynamics in catalytic nanoparticles have been significantly impacted by the challenges associated with high degrees of noise \citep{lawrence2019, levin2020, lawrence2021, vincent2021}.  Moreover, the large number of noisy image frames required to fully map out the details of the spatio-temporal behavior requires the collection of large image data sets---on the order of terabytes \citep{lawrence2019}. As noted in \cite{lawrence2019}, there is a continuing need for algorithms to automate the extraction of relevant features from images to facilitate the assessment of dynamics, and to do so in the presence of an immense amount of noise. Denoising methods based on convolutional neural networks trained on simulated nanoparticle configuration images have been developed specifically to deal with such ultra-noisy nanoparticle videos \citep{mohan2020}. However, it may not always be feasible to implement these methods in practice, frame-by-frame, on the aforementioned terabytes of data.


In this article, we present an algorithm for detecting features in, and hypothesis testing of, severely noisy images based off of topological data analysis (TDA)---more specifically, cubical persistent homology (cPH) \citep{edelsbrunner2002, kaczynski2006, mischaikow2013}. Cubical homology is more naturally suited to imaging as it treats images and their connectivity in a natural manner, with no need for triangulation of the inherently pixelated data \citep{kaczynski2006}. Indeed, cubical persistent homology \citep{garin2019, rieck2020, lawson2021, chung2018} has been applied to great effect in the statistics, machine learning, and imaging communities in recent years. Additionally, the algorithms which compute cubical persistent homology are often much faster than their simplicial counterparts \citep{wagner2012}. We will discuss the background beyond these concepts and provide intuition for them in Section~\ref{s:cubical}. 

Topological methods as used in this paper have the advantage over traditional methods in materials science in that they are isometry invariant. Cubical homology in particular is invariant to translations, thus providing robustness against minor perturbations of atomic columns and ridges across frames. In the previous few years, Applications of methods in TDA to materials science (beyond cPH) have seen greater adoption. In \citet{motta2018}, the authors use functionals of persistent homology, such as the variance of $H_0$ barcode lengths and the sum of $H_1$ barcode lengths, to characterize the order of a nearly hexagonal planar lattice---e.g., a perturbed Bravais lattice. In cluster physics, \citet{chen2020} examined the ability of topological features, in conjunction with machine learning, to assess and predict ground-state structure-energy relationships in lithium clusters. In a similar context, \cite{jiang2021} examined a topological invariant called ``atom-specific persistent homology'' to predict formation energy of crystal structures. \citet{nakamura2015} used the persistence diagram to characterize medium-range order in amorphous materials. This is all to say that employing TDA in material science applications is a fruitful endeavor. 

It is useful to note that there have been applications of TDA in image segmentation problems \citep{vandaele2020, chung2018}. We believe this is a fruitful avenue to pursue, as binarized images are conceptually simpler. We provide a brief illustration of the ability of cPH to recover underlying topological structure by applying PD thresholding \cite{chung2018} to threshold and binarize our images. 

Here we use methods from TDA to classify the fluxional behavior of nanoparticles. As alluded to earlier, we identify atomic columns (see Figure~\ref{f:evo3x3}) via their estimated \emph{persistence}---quantified within each individual frame as the difference in greyscale threshold at which a given dark region appears and when it merges with another dark region that appeared before it. Such a paradigm is called the \emph{elder rule} and is described in detail in \cite{edelsbrunner2010}. The appearance of a connected component in our images corresponds to the appearance of a local minima. Though methods such as  \cite{mpfit2020, atomap} use the local minima of images as initial locations for fitted Gaussians (from which the intensity is estimated), here we estimate the intensity by the ``lifetimes'' of the local minima, i.e. the concept of persistence that we defined at the start of this paragraph. Thus, the lifetime of a local minimum (its persistence) is defined as the difference in pixel values between the appearance of the local minimum and the pixel value at which it merges with another longer-lived local minimum. 

Recently, there have been forays into using the statistics that appear in TDA in hypothesis and goodness-of-fit testing \citep{biscio_acf, blumberg2014, fasy2014, robinson2017, cericola2017, vejdemo2020}. The present article is the first to attempt this in a cubical setting and the first to evaluate the efficacy of certain topological summaries to capture relevant topological features in the presence of powerful noise. In particular, both \emph{persistent entropy} \citep{pers_entropy, rucco2016} and the \emph{accumulated lifetime persistent survival} (ALPS) statistic---a new topological summary---perform well and evince good statistical power. It is our opinion that they would work well in a litany of tasks in summarizing noisy videos, particularly the ALPS statistic when the number of features and their intensity are both salient. On a final note, we prove a result demonstrating the false discovery rate of a certain multiple Monte Carlo test tends to any $\alpha$ almost surely. This yields a theoretically sound, as well as computationally efficient, means of multiple testing in persistent homology, improving on previous studies in the area \citep{cericola2017, vejdemo2020}. 

It is our hope that the algorithm and hypothesis testing framework we have devised here can provide an off-the-shelf method of statistical detection of atomic structure in a flexible manner for those in the material science community and others that deal with necessarily noisy images. Our method performs well in the standard nanoparticle imaging task of determining position and location of atomic columns \citep{atomap, levin2020} and also performs well against the state-of-the-art \citep{xu2022}. Additionally, these topological methods do not presume a particular structure to the image data.  For example, Gaussian peak fitting and blob detection (ibid.) both assume an elliptical structure to features in the image which are present when individual columns of atoms are well aligned. However, in many cases, the crystal may be tilted and individual columns may not be resolved but planes of atoms may be visible. Moreover, during structural dynamics and in the presence of high concentrations of crystal defect, the structure of the image intensity may be complicated and rapidly changing. Thus to elucidate structural dynamics, it is important to have an image analysis method that can adapt to the changing structure of the image contrast and does not presuppose a particular image form. Finally, between the persistence entropy and the ALPS statistic, we offer a choice in how conservative a practitioner wants to be in determining which atomic features are statistically significant.


In the following, we will discuss the cubical persistence algorithm we devised to process the extremely noisy videos at high time resolution in Section~\ref{s:top_sum_nano}, along with the ability of said algorithm to recover atomic features in simulated datasets before and after the application of noise in Section~\ref{s:noise_experiment}. The main statistical contribution is in Section~\ref{s:sdht}, wherein we describe the parametric assumptions of the noise region of the data, check those assumptions and investigate various topological summaries of persistence diagrams for our Monte Carlo goodness-of-fit test. Having a means of testing whether or not there is noise in the frame of a video has great utility when the presence of noise nearly overwhelms all of the signal, as is the case here. 

Before continuing to the description of our algorithm and the rest of the paper, we must first introduce and detail the concepts of cubical homology and persistence. 

\section{Background} \label{s:cubical}


\subsection{Cubical sets} 
\begin{figure}
\centering
\includegraphics[width=4.5in]{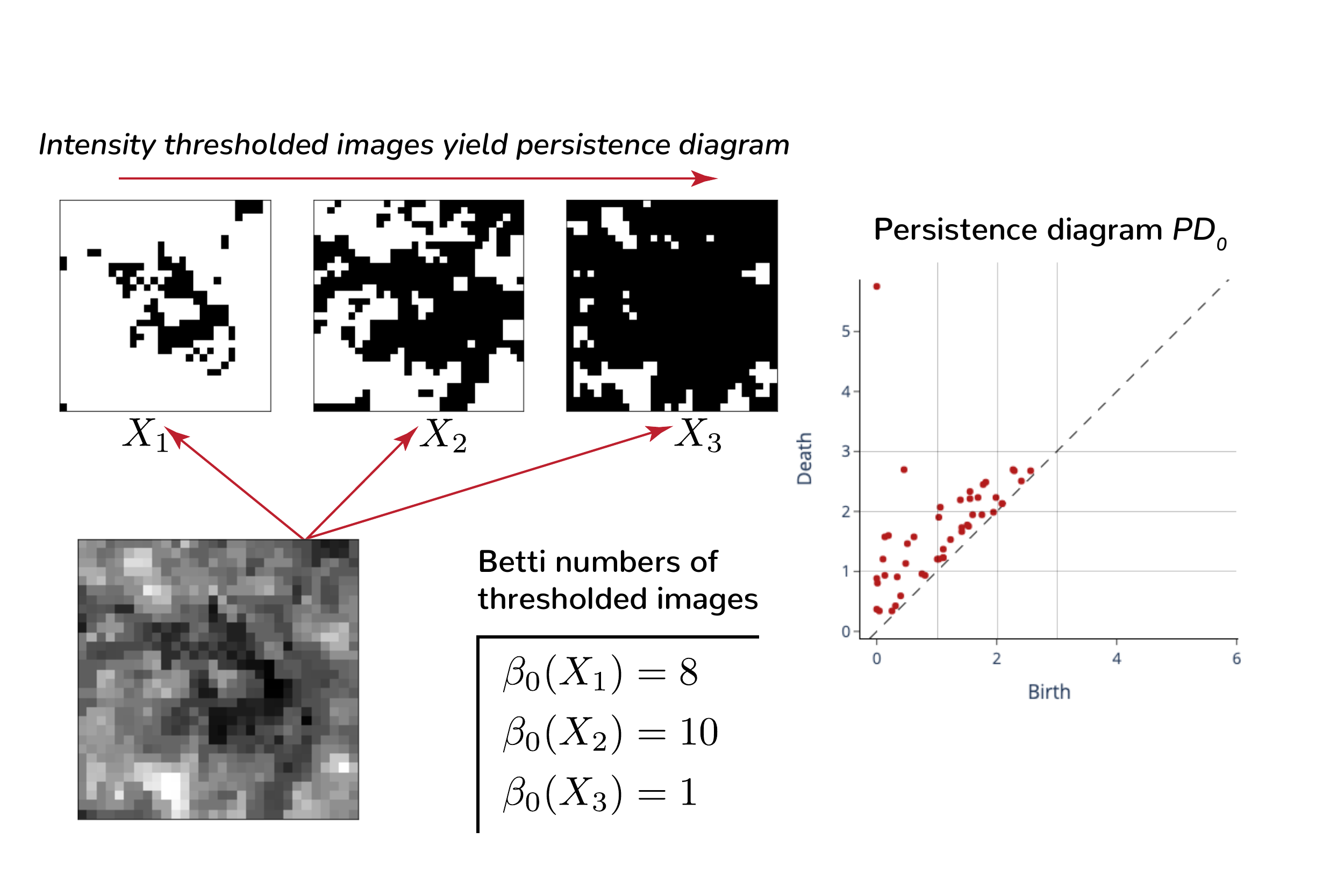}
\caption{Thresholded images $X_t$ with pixel intensities at most $t$, of a single atomic column viewed at three pixel intensity thresholds. The greyscale thresholds for $X_1, X_2, X_3$ equal 1, 2 and 3, respectively. The persistence diagram $PD_0$ of the atomic column in the bottom left can be seen to the right. The persistence diagram to the right depicts the pixel values at which a given black-connected region---a feature in $H_0$---appears (i.e. is ``born'') and when it merges or ``dies''. Betti numbers at the three greyscale thresholds can be seen in the table in the lower right.}
\label{f:evo3x3}
\end{figure}

The tool that we use to assess shapes in images in this paper is cubical persistent homology (cPH). One reason for considering a cubical representation of an image is that it is the most natural construction for a 2-dimensional image from the perspective of topology, as detailed in \citet{kovalevsky1989}. Another reason is that cubical persistence algorithms run in linear time in the number of pixels $n$ in the image $\mathcal{I}$, whereas traditional approaches to persistent homology, such as using the \v{C}ech or Vietoris-Rips complex, can only be computed in polynomial time in the number of points in the point cloud \citep{wagner2012}. To introduce cPH we must introduce the notion of cubical homology and the objects it acts on: cubical sets. The cubical sets we consider here are collections of unit squares (\emph{2-dimensional elementary cubes}) of the form 
\[
[i, i+1] \times [j, j+1],
\]
along with all intervals (\emph{1-dimensional elementary cubes}) and vertices (\emph{0-dimensional elementary cubes}) on the boundaries, where $i$ and $j$ are integers---i.e. $(i,j) \in \mathbb{Z}^2$. Once we have a cubical set (or \emph{cubical complex}) $X$, we can calculate \emph{homology}.  Loosely speaking, homology is an algebraic method of assessing ``connectivity''  in various dimensions. Given a cubical set $X$, we can associate a homology group\footnote{In this paper, we assume these are vector spaces over the field on two elements $\mathbb{Z}_2 = \{0,1\}$} over $H_k(X)$---which captures $k$-dimensional shape information---for each nonnegative integer $k$. Of great interest are the dimensions of these homology groups, which are called the \emph{Betti numbers} of $X$ and are denoted $\beta_k(X)$, or $\beta_k$ when the underlying cubical set is clear from the context. For example, the $0^{th}$ Betti numbers $\beta_0$ represent connected components and $\beta_1$ represents loops/holes. As the relevant features we aim to capture are darker than their surroundings, we will henceforth focus on $\beta_0$. One can see Figure~\ref{f:evo3x3} for the calculation of $\beta_0$ at various greyscale thresholds. Note that in Figure~\ref{f:evo3x3} the black pixels are the ones included in our cubical set---in this sense all cubical sets/complexes that we treat here can be considered as binary images. For more information on cubical homology, one may refer to Chapter 2 of \citet{kaczynski2006}.

\begin{figure}
    \centering
    \includegraphics[width=4.5in]{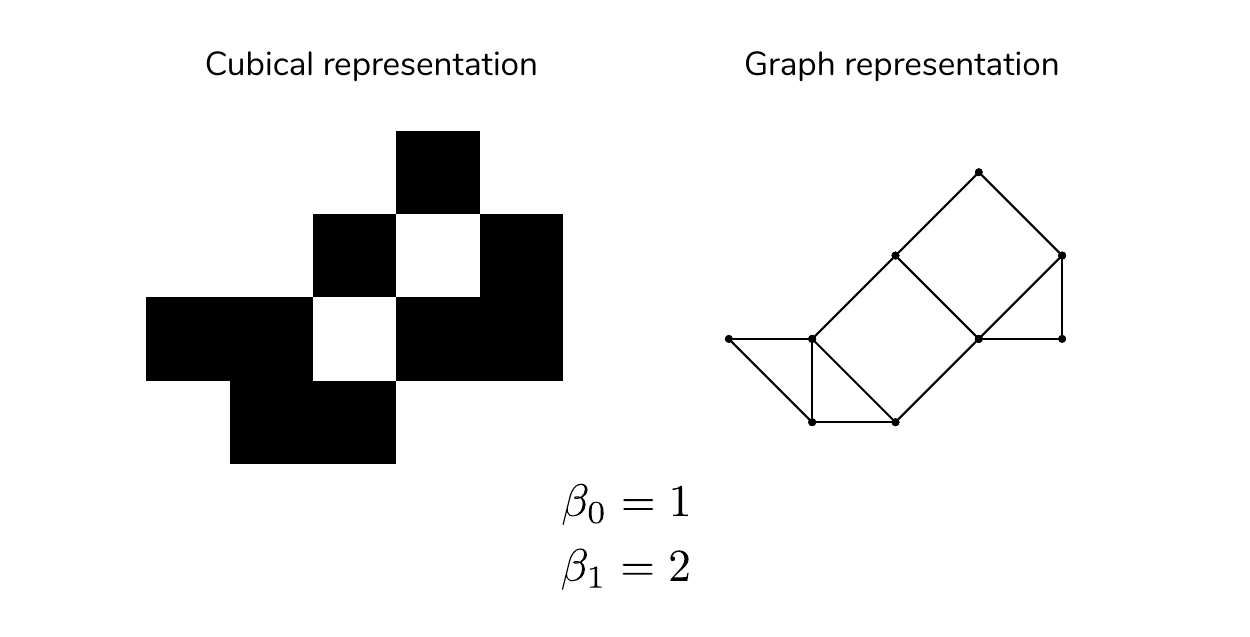}
    \caption{Connectivity information in the cubical set to the left can is conveyed by the graph to the right. Note that there is one connected component and two ``holes'', because the black region connects via the diagonals and the white region does not. }
    \label{f:exa_cub_graph}
\end{figure}

\subsection{Image model} In this article, a (2-dimensional) \emph{image map} is a function $I: \mathbb{Z}^2 \to [0,1]$, where $I(p) = 0$ indicates that $p$ is a black pixel and $I(p)=1$ indicates that $p$ is a white pixel. We call the smallest rectangle $[k, k+m] \times [l, l+n] \subset \R^2$ which contains all the black pixels---i.e. on which $1-I > 0$---the \emph{image set}, which we denote $\mathcal{I}$. Via appropriate normalization, every image that is bounded or has a finite image set can be modified to have pixels between 0 and 1. We may choose to set the codomain of $I$ to $\R$ (or the first $n$ nonnegative integers) instead. As previously mentioned, for the purposes of cubical homology and cubical persistence we must identify $I$ in some appropriate way with a collection of cubical sets. We do this here by the construction of another function $I'$ on the family $\{[i, i+1] \times [j, j+1]: (i,j) \in \mathbb{Z}^2\}$ of unit squares with integer vertices. For any such $\tau = [i, i+1] \times [j, j+1]$ we define our filtration function
\[
I'(\tau) := I(i,j).
\]
For lower dimensional elementary cubes $\tau$, such as intervals or vertices, we define the value of $I'$ to be the minimum $I(i,j)$ such that $\tau \subset [i, i+1] \times [j, j+1]$. This is consistent with the definition used in the persistent homology software \texttt{GUDHI} Python library, which we use for our calculations throughout this article \citep{gudhi_cubical}. We consider the homology of sublevel set filtrations\footnote{Other filtrations could be considered here; however, besides the sublevel set filtration, all require choosing a threshold at which to binarize the image \citep{turkes2021, garin2019}--see also the opening paragraph of Section~\ref{s:noise_experiment}.}, so darker pixel values will appear first. The cubical complex construction we use here, treating pixels as unit squares (i.e. top-dimensional) is also known as the $T$-construction and it is dual in some sense to treating pixels as points (or vertices), rather than unit squares \citep{garin2020duality}. 

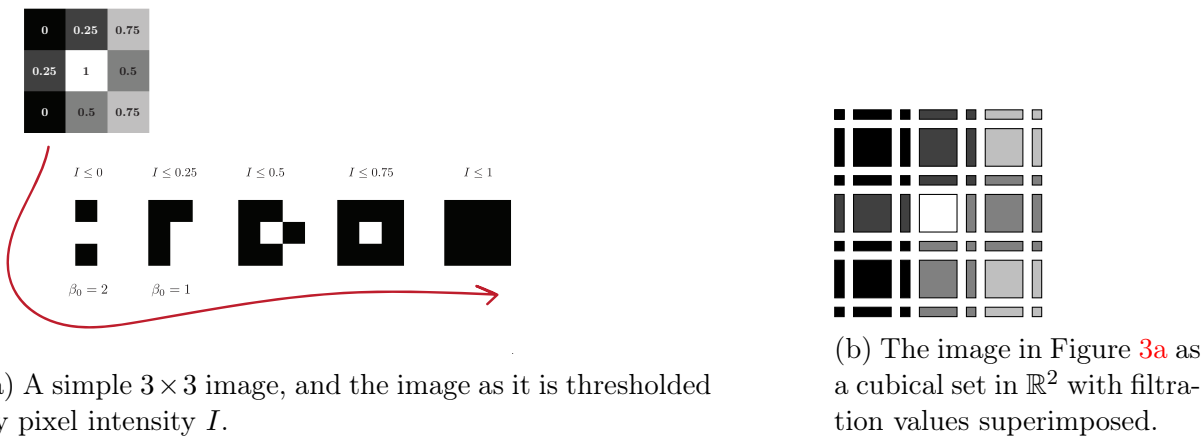
\begin{figure}[t]
\begin{subfigure}[b]{0.6\textwidth}
\includegraphics[width=3in]{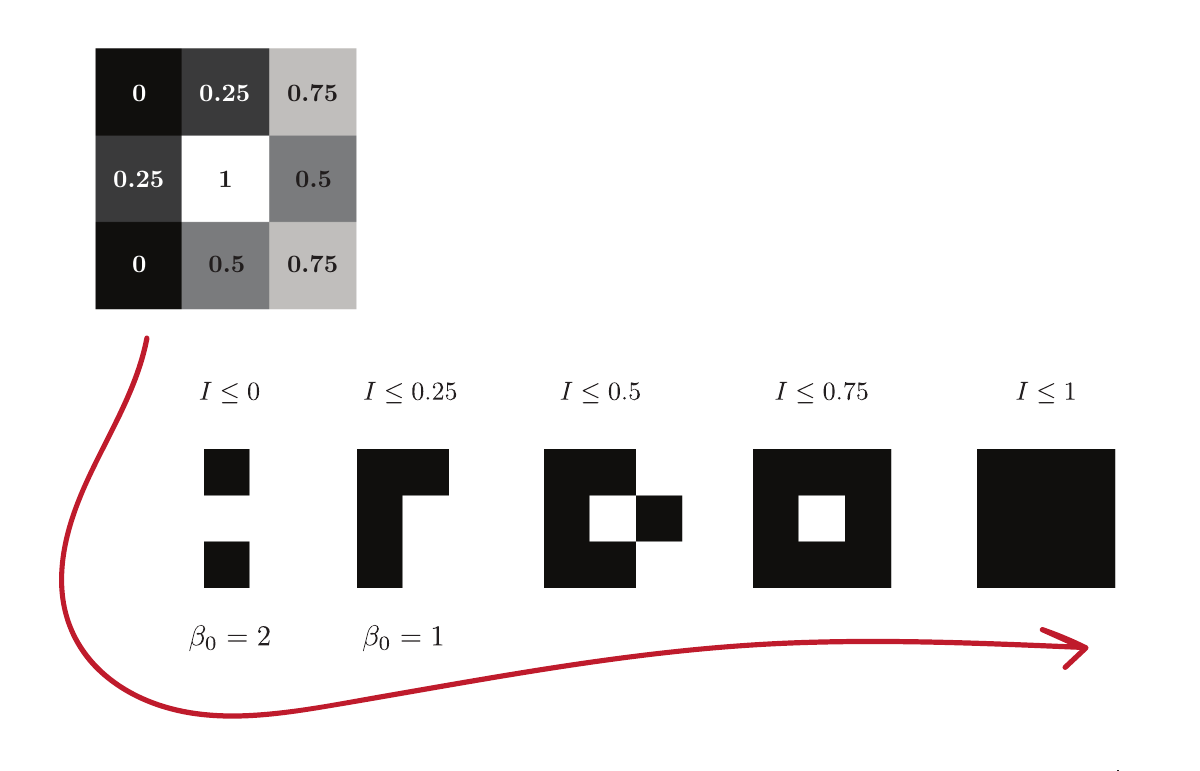}
\caption{A simple $3 \times 3$ image, and the image as it is thresholded by pixel intensity $I$.}
\label{f:demon_cube}
\end{subfigure}
\hfill
\begin{subfigure}[b]{0.3\textwidth}
\begin{tikzpicture}[scale=1.25]
\draw [fill=black] (0,0) rectangle (0.1, 0.1);
\draw [fill=black] (0.2,0) rectangle (0.6, 0.1);
\draw [fill=black] (0.7,0) rectangle (0.8, 0.1);
\draw [fill=gray2] (0.9,0) rectangle (1.3, 0.1);
\draw [fill=gray2] (1.4,0) rectangle (1.5, 0.1);
\draw [fill=gray3] (1.6,0) rectangle (2, 0.1);
\draw [fill=gray3] (2.1,0) rectangle (2.2, 0.1);

\draw [fill=black] (0,2.1) rectangle (0.1, 2.2);
\draw [fill=black] (0.2,2.1) rectangle (0.6, 2.2);
\draw [fill=black] (0.7,2.1) rectangle (0.8, 2.2);
\draw [fill=gray1] (0.9,2.1) rectangle (1.3, 2.2);
\draw [fill=gray1] (1.4,2.1) rectangle (1.5, 2.2);
\draw [fill=gray3] (1.6,2.1) rectangle (2, 2.2);

\draw [fill=black] (0,0.2) rectangle (0.1, 0.6);
\draw [fill=black] (0,0.7) rectangle (0.1, 0.8);
\draw [fill=gray1] (0,0.9) rectangle (0.1, 1.3);
\draw [fill=black] (0,1.4) rectangle (0.1, 1.5);
\draw [fill=black] (0,1.6) rectangle (0.1, 2.0);
\draw [fill=black] (0,2.1) rectangle (0.1, 2.2);

\draw [fill=gray3] (2.1,0.2) rectangle (2.2, 0.6);
\draw [fill=gray2] (2.1,0.7) rectangle (2.2, 0.8);
\draw [fill=gray2] (2.1,0.9) rectangle (2.2, 1.3);
\draw [fill=gray2] (2.1,1.4) rectangle (2.2, 1.5);
\draw [fill=gray3] (2.1,1.6) rectangle (2.2, 2.0);
\draw [fill=gray3] (2.1,2.1) rectangle (2.2, 2.2);

\draw [fill=black] (0.7, 0.2) rectangle (0.8, 0.6);
\draw [fill=black] (0.7, 0.7) rectangle (0.8, 0.8);
\draw [fill=gray1] (0.7, 0.9) rectangle (0.8, 1.3);
\draw [fill=black] (0.7, 1.4) rectangle (0.8, 1.5);
\draw [fill=black] (0.7, 1.6) rectangle (0.8, 2.0);

\draw [fill=gray2] (1.4, 0.2) rectangle (1.5, 0.6);
\draw [fill=gray2] (1.4, 0.7) rectangle (1.5, 0.8);
\draw [fill=gray2] (1.4, 0.9) rectangle (1.5, 1.3);
\draw [fill=gray1] (1.4, 1.4) rectangle (1.5, 1.5);
\draw [fill=gray1] (1.4, 1.6) rectangle (1.5, 2.0);

\draw [fill=black] (0.2,0.7) rectangle (0.6, 0.8);
\draw [fill=black] (0.2,1.4) rectangle (0.6, 1.5);

\draw [fill=gray2] (0.9,0.7) rectangle (1.3, 0.8);
\draw [fill=gray1] (0.9,1.4) rectangle (1.3, 1.5);

\draw [fill=gray2] (1.6,0.7) rectangle (2, 0.8);
\draw [fill=gray2] (1.6,1.4) rectangle (2, 1.5);


\draw [fill=black] (0.2,0.2) rectangle (0.6, 0.6);
\draw [fill=gray2] (0.9,0.2) rectangle (1.3, 0.6);
\draw [fill=gray3] (1.6,0.2) rectangle (2, 0.6);

\draw [fill=gray1] (0.2,0.9) rectangle (0.6, 1.3);
\draw [fill=white] (0.9,0.9) rectangle (1.3, 1.3);
\draw [fill=gray2] (1.6,0.9) rectangle (2, 1.3);

\draw [fill=black] (0.2,1.6) rectangle (0.6, 2);
\draw [fill=gray1] (0.9,1.6) rectangle (1.3, 2);
\draw [fill=gray3] (1.6,1.6) rectangle (2, 2);

\end{tikzpicture}



\caption{The image in Figure~\ref{f:demon_cube} as a cubical set in $\R^2$ with filtration values superimposed.}
\end{subfigure}
\caption{Thresholded black and white images in a filtration (left) and the corresponding cubical set of said image (right). Note the appearance of a hole at $I(p) = 0.5$ and the death of the hole entering $I(p) = 1$.}
\end{figure}

\begin{remark}
An important heuristic below (see Figures~\ref{f:evo3x3}, \ref{f:exa_cub_graph}, and \ref{f:demon_cube}) is that \emph{black} pixels ``connect'' via the diagonals. A connected component in cubical homology (contributing to $\beta_0$) is a \emph{8-connected black region of pixels bordered by either white pixels or the edge of the image}\footnote{Recall that all pixels not in the image (image set) are de facto white pixels in terms of cubical homology.}. In terms of chess, a connected component in our construction is a region that can be traversed via queen moves. The equivalence between the notion of connectivity for the $T$-construction and $8$-connectedness was established in \cite{kovalevsky1989}. 


\end{remark}

\subsection{Persistent homology}

Suppose now that we have the collection of cubical complexes $\X = \{X_t\}_{t \in [0,1]}$, where 
\[
X_t := \bigcup_{(i,j) \in I^{-1}([0,t])} [i, i+1] \times [j, j+1],
\] 
or alternatively, $X_t =  I'^{-1}([0,t]).$ It is clear that for $s \leq t$ we have $X_s \subset X_t$ and thus $\X = \{X_t\}_{t \in [0,1]}$ defines a \emph{filtration} of cubical complexes. Given the inclusion maps $\iota_{s,t}$, for $s \leq t$ there exist linear maps between all homology groups
\[
f^{s,t}_k: H_k(X_s) \to H_k(X_t),
\]
which are induced by $\iota_{s,t}$ \citep[see chapter 4 of][]{kaczynski2006}. The \emph{persistent homology groups} of the filtered image $\X$ are the quotient vector spaces $\im\, f^{s,t}_k$ whose elements represent shape features---such as connected components or holes---called \emph{cycles} that are ``born'' in or before $X_s$ and that ``die'' after $X_t$. The dimensions of these vector spaces are the \emph{persistent Betti numbers} $\beta_k^{s,t}$. Heuristically, a cycle\footnote{Technically speaking these are equivalence classes of cycles, which are equivalent modulo a boundary.} $\gamma \in H_k(X_s)$ is born at $X_s$ if it appears for the first time in $H_k(X_s)$---formally, $\gamma \not \in H_k(X_r)$, for $r < s$. The cycle $\gamma \in H_k(X_s)$ dies entering $X_t$ if it merges with an older cycle (born at or before $s$) entering $H_k(X_t)$. The $k^{th}$ persistent homology of $\X$, denoted $PH_k$, is the collection of homology groups $H_k(X_t)$ and maps $f_{s,t}^k$, for $0 \leq s \leq t \leq 1$. All of the information in the persistent homology groups is contained in a multiset in $\R^2$ called the \emph{persistence diagram} \citep{edelsbrunner2010}. 

The $k^{th}$ persistence diagram of $\X$, denoted $PD_k$, consists of the points $(b, d)$ with multiplicity equal to the number of the cycles that are born at $X_b$ and die entering $X_d$. Often, the diagonal $y = x$ is added to this diagram, but we need not consider this here. See Figure~\ref{f:evo3x3} for an illustration of the $0^{th}$ persistence diagram associated to a filtration of a given greyscale image. For this study, we focus on $PD_0$. In this particular setup, if $(b,d) \in PD_0$, this indicates there is a local minimum of the image $\X$ at some pixel $p^+$ with $I'(p^+) = b$ and $d$ represents the greyscale threshold at which the connected component containing $p$ merges with a connected component containing a local minimum with birth time \emph{less than} $b$. In this case, $p^+$ is called a \emph{positive cell} and gives birth to a connected component in $PH_0$. Furthermore, we can also find an interval $\tau^-$ that kills such a feature, i.e. $I'(\tau^-) = d$ \citep[cf.][]{geom_top2018}.




%
%

\section{The algorithm} \label{s:top_sum_nano}

\subsection{Description} 

In this section, we describe our algorithm for extracting shape, location, and intensity information from ultra-noisy images. To speed computation we may restrict our attention to a rectangular subimage $\mathcal{L} \subset \mathcal{I}$---see Figure~\ref{f:sub_ill} for a depiction of this process. Let us denote the restriction of $I$ to $\mathcal{L}$ by $\IL$. Hence, we process our subimage $p \mapsto I(p)$ according to the following steps:

\begin{enumerate}
	\item Identify polygonal\footnote{For specifying polygonal regions and which pixels are contained in them, we use the \texttt{Shapely} Python library \citep{shapely}.} nanoparticle region $R \subset \R^2$, which we will use to exclude pixels that lie outside of $R$. 
	\item Smooth the image with a Gaussian filter, with smoothing parameter $\sigma$.
	\item Compute $PH_0$ for image $\IL$ , based off of the filtration function $\IL'$. Note that one should have $R \subset \mathcal{L}$. 
	\item If the pixel $p^+$ associated to the creation of connected component is located outside of $R$, remove point associated to $p^+$ from $PD_0$. 
	\item (Optional) Remove features with persistence at or below a threshold $\eta \geq 0$ from $PD_0$.
\end{enumerate}

For our image $I$, we denote the output of this algorithm as $A(I)$, which consists of the locations of atomic columns (or other atomic features) as well as their persistence (or, intensity). As such, $A(I)$ may be considered as a finite subset of $R \times [0, \infty)$. Equivalently, we may consider $A(I)$ as a marked point process on $R$ with mark space $[0, \infty)$, as we assumed that our image $I$ is subject to noise. Additionally, we denote the thresholded output as $A_\eta(I)$, so that the original output may be considered as $A(I) = A_0(I)$. Formally $A_{\eta}(I) = \{ (p, l) \in A(I): l > \eta\}$, where $l = d-b$ are the \emph{lifetimes} associated to the pixels $p^+$. Note that if we preprocess by restricting our image to $\mathcal{L}$, the algorithm requires only $R$ and $\sigma$ to be specified. 

With this in tow, let us now examine the algorithm in greater detail. For step 1, as cubical persistence does not consider the size of connected component per se, we remove image features (corresponding to atomic columns in our application) outside of some region, which we know either corresponds to noise or the structure of which is of no interest to us. Now to smooth the image, we consider the image $I_\sigma := G_\sigma * I$, convolved with the spherical Gaussian kernel $G_\sigma$ where 
\[
I_\sigma(i,j) = (G_\sigma * I)(i,j) = \frac{1}{C} \sum_{(k, l) \in \mathbb{Z}^2} G_\sigma(i - k, j- l)I(k,l),
\]
and 
\[
G_\sigma(x,y) := \frac{1}{2\pi\sigma^2}e^{-\frac{x^2+y^2}{2\sigma^2}},
\]
where $C = \sum_{(x, y) \in \mathbb{Z}^2} G_\sigma(x,y)$. By convention let us take $G_0 = \delta_{(0,0)}$ be the Dirac delta function at the origin, i.e. $\delta_{(0,0)}(x,y) = 1$ if and only if $x=y=0$. 

\subsection{Justification of the Gaussian kernel} With respect to a suitably large class of kernels and signals, the Gaussian kernel is the only kernel that ``preserves'' local minima in continuous 1-dimensional signals as $\sigma$ increases \citep{babaud1986}. This occurs in the sense that the value of a signal at local minima increases as smoothing increases---local minima can only be destroyed and not created. This is particularly relevant because the locations of the pixels which create connected components---that correspond to locations of atomic columns---represent local minima \citep{robins2011}. Such stability may not strictly be the case for 2-dimensional discrete signals for low-levels of smoothing \citep{lindeberg1990}. However, when $\sigma \geq 1$ (as is the case for all practical settings in this article), the ideal ``Lindeberg'' kernel (which does not increase local minima) and the discrete Gaussian kernel coincide to a large degree \citep{getreuer2013}. 

Furthermore, the discretized Gaussian kernel does not increase the number of local minima from the unsmoothed image \citep{lindeberg1990}. Therefore, in theory, tuning the $\sigma$ parameter appropriately will allow for cubical persistence to recover the precise number of relevant atomic columns, their location as well as an estimate of their intensity. Additionally, the Gaussian kernel has been empirically shown to introduce the fewest ``image artifacts'' \citep{levin2020}. Another option would be to convolve our image with an elliptical Gaussian kernel, such as in \cite{kong_blob} or to apply our cubical homology algorithm directly to a scale-space representation \citep{lindeberg1998}. Decay rates of the norms of persistence diagrams after convolution with a Gaussian kernel have also been established \citep{chen2011diffusion}.

We now compare the performance of the above algorithm with other methods of finding atomic column positions and show how it performs better in certain cases and works well as a method for finding initial positions of atomic features.



%

\begin{figure}[!ht]
\includegraphics[width=4.8in]{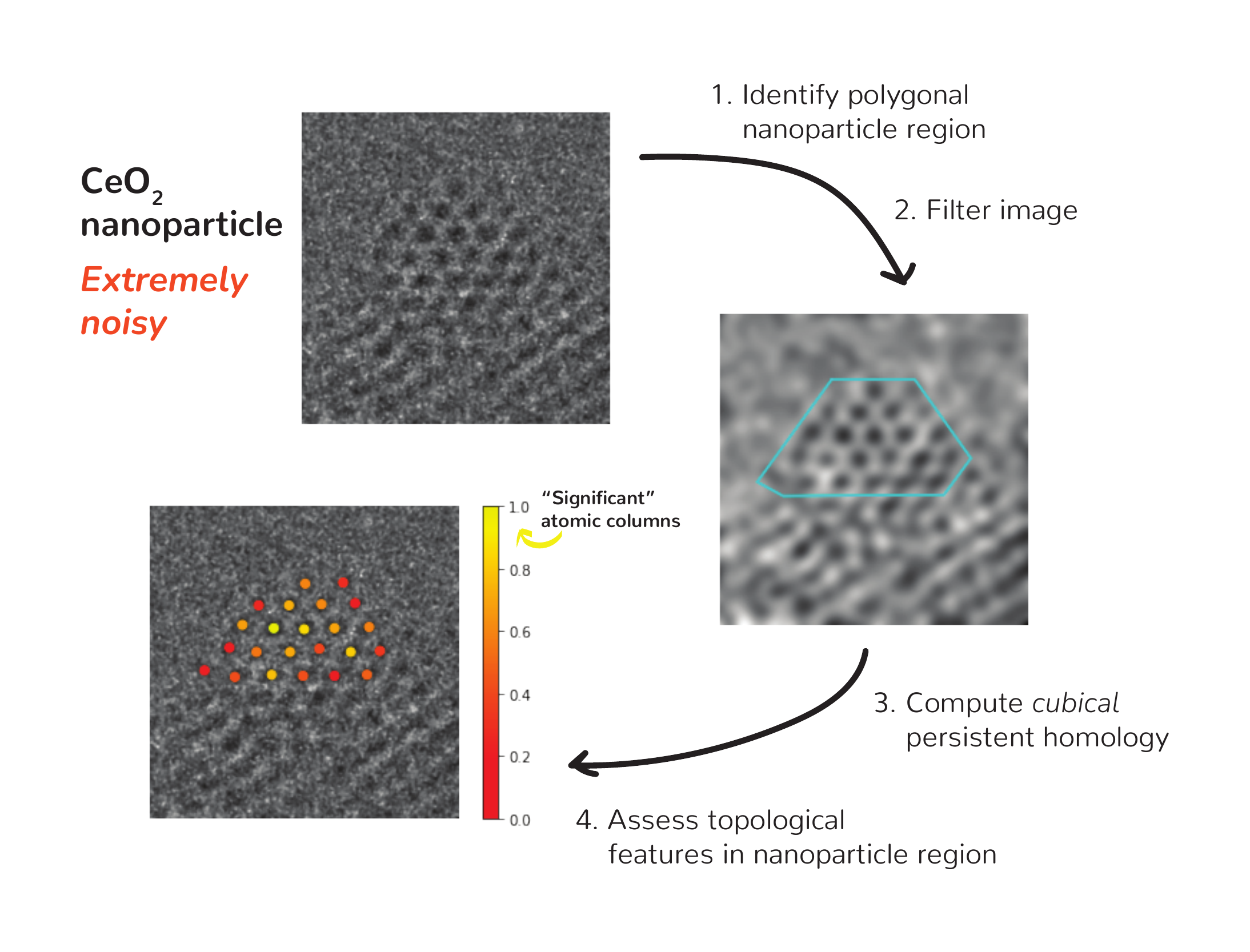} 
\caption{Illustration of atomic column processing pipeline, applied to a cerium oxide ($\text{CeO}_2$) nanoparticle as described in \citet{lawrence2019}. Persistence in colorbar measured as proportion of longest barcode.}
\label{f:atomcol_pipe}
\end{figure}

%

\section{Noise experiment}\label{s:noise_experiment}

In this section, we assess the ability of the algorithm we introduced in Section~\ref{s:top_sum_nano} to perform as well as the combined blob detection/gaussian peak fitting (GPF) method from \citep{xu2022} in recovering the number, location, and intensity of the atomic columns in a nanoparticle image corrupted by Poisson noise. 
Images were simulated and Poisson noise was added according to the method described in Section 2.4 of \citet{xu2022}. Persistent homology has purported to be robust to noise. In \cite{skraba2021wasserstein}, the authors offer a cubical version of the classical persistence stability theorem \citep{cohenstein_stab}, stating that if two image maps are close, then their persistence diagrams are close as well. There have been other efforts to quantify experimentally the noise robustness in cPH. In \citet{turkes2021}, the authors demonstrated the empirical robustness of sublevel set (greyscale) filtrations for cPH under affine transformations and additive noise (such as the Poisson noise encountered in our application). Observation of their results buttresses our argument that pre-smoothing and thresholding an image can faithfully recover the underlying topology. 

Here, we assess the mean and standard error of three statistics related to the recovery of the homology of simulated nanoparticle images based on the output of our algorithm. Let us denote our smoothed noisy \emph{simulated} images as $I^s_{\sigma, j}$, $j = 1, \dots, 10$. Let $I_{\sigma}$ be the $\sigma$-smoothed version of the observed image $I$. We assess the ability of our cubical homology algorithm applied to the noisy images, to recover atomic column position and intensity that the algorithm outputs on the noise-free image. We assess the number of columns output by the algorithm; the mean (Pearson) correlation $\hat{\rho}(I_{\sigma}, I^s_{\sigma, j})$ of the intensity of the derived columns in the noisy output $A(I^s_{\sigma, j})$ compared to the output of the noise-free image $A(I_\sigma)$; and the Hausdorff distance 
\[
d_H(I_{\sigma}, I^s_{\sigma, j}) := \max \Bigg\{ \max_{(x,l) \in A(I_\sigma)} \min_{(y,l_j) \in A(I^s_{\sigma, j})} \norm{x-y}, \max_{(y,l_j) \in A(I^s_{\sigma, j})} \min_{(x,l) \in A(I_\sigma)}  \norm{y-x} \Bigg\}
\]
between the locations of the columns in the algorithms output. In this section, if the death time equals $d=\infty$, we set $d$ to be the largest finite death time in the persistence diagram. If we chose $d$ to be the largest pixel value in the image, the death time for the longest barcode is a significant outlier. To assess the performance of each algorithm $A$ (i.e. cubical homology vs. blob detection/GPF), we assess the output of $A$ with the same parameter set $\Theta$ (such as $\Theta = \{\sigma\}$, using our algorithm) on both the noisy and noise-free simulated images. For the combined blob detection/GPF method\footnote{The locations of the blobs/atomic columns was initially calculated using the \texttt{blob\_log} function the Python \texttt{skimage} library, as in \cite{xu2022}. The algorithm was applied in the same fashion as \cite{xu2022} to ensure optimality of parameters chosen and a fair comparison of the methods.}, the mean correlation was 0.9816 with standard error 0.0046 and the mean Hausdorff distance was 2.319 with standard error 1.088. The corresponding results for our cubical homology algorithm can be seen in Table~\ref{t:noise_exp_res}. 
\begin{figure} 
\includegraphics[width=3.9in]{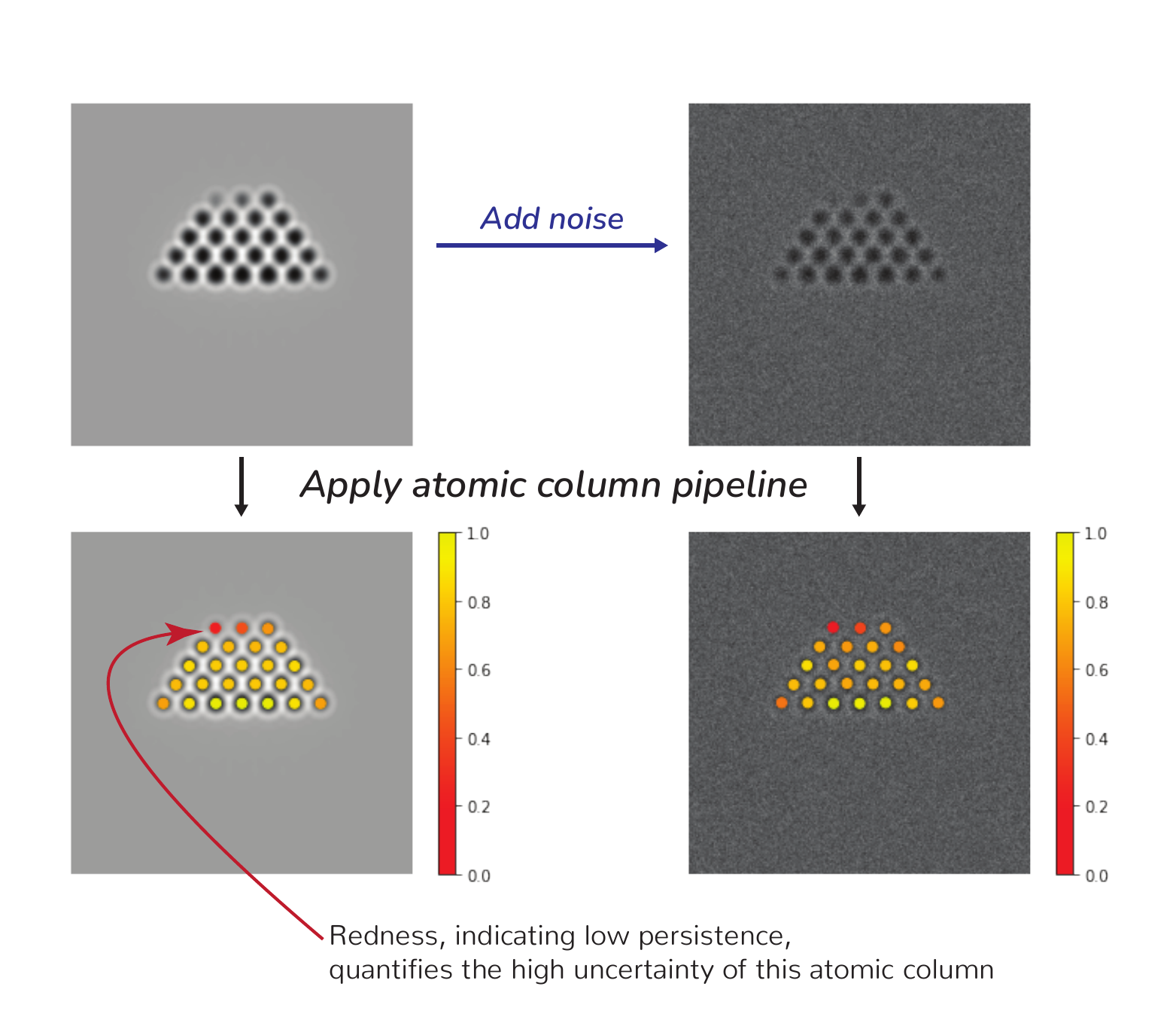} 
\caption{Recovery of the $0^{th}$ homology (i.e., atomic columns) in the presence of noise. Note that persistence in these images is measured in terms of the proportion of longest barcode.}
\label{f:noise_recovery}
\end{figure}

\begin{table}[!ht]
\bgroup
\def\arraystretch{1.25}
\begin{tabularx}{0.99\textwidth}{| p{1.3in} | X | X | X | X | X | X | X |}
\hline
$\sigma$ & 0 & 2 & 4 & 6 & 8 & 10 & 12 \\
\hline
$N(I^s_{\sigma, j})$ \newline \footnotesize{\#Columns recovered} & 3055 \newline (22.08) & 55 \newline (5.02) & 25.2 \newline (0.4) & \textbf{25 \newline (0)} & \textbf{25\newline (0)} & \textbf{25 \newline (0)} & 24.5 \newline (0.5) \\
\hline
$d_H\big(I_{\sigma}, I^s_{\sigma, j}\big)$ \newline \footnotesize{Hausdorff distance} & 26.61 \newline (0.35) & 23.82 \newline (1.02) & 5.16 \newline (5.68) & \textbf{1.74 \newline (0.88)} & \textbf{2.21 \newline (1.15)} & \textbf{4.19 \newline (2.39)} & 21.52 \newline (17.26)  \\
\hline
$\hat{\rho}(I_{\sigma}, I^s_{\sigma, j})$ \newline \footnotesize{Pearson correlation} & N/A & N/A & $0.961^*$ \newline (0.016) & \textbf{0.962 \newline (0.023)} & \textbf{0.929 \newline (0.052)} & \textbf{0.843\newline (0.116)} & $0.721^*$ \newline (0.212)  \\
\hline
\end{tabularx}
\egroup
\vspace{12pt}
\caption{In each cell is the mean (standard error) of the summary described in the row label for the smoothing parameter $\sigma$ seen in the column label, across all 10 noisy frames. An asterisk (*) means that mean/standard error was only taken over the (less than 10) frames where all 25 columns were recovered---respectively 8 and 5 out of 10 instances for $\sigma=4$ and $\sigma=12$. N/A indicate that in none of the 10 frames were the correct number of columns identified. The Hausdorff distance is measured here in pixels.} 
\label{t:noise_exp_res}
\end{table}

For appropriate $\sigma$ the mean Hausdorff distance was much better using our method, though this could be attributed to the fact that a variety of $\sigma$ smoothing values (50 different values from 6 to 9) were used to find the best blobs in blob detection, whereas $\sigma$ remained fixed for the ground truth image as well as the noisy image in our method. As one can see, the mean correlation performs similarly to blob detection, however for lower values of $\sigma$, the actual intensities of the 25 true atomic columns is retained to a much higher degree, owing to less influence from surrounding pixels attenuating their signal. Setting the threshold $\eta$ such that we choose the 25 largest persistence values, we achieve correlations of $0.893\ (0.042)$ when $\sigma=2$ and $0.961\ (0.015)$ when $\sigma=4$ between the derived intensities of $I_{\sigma}$ and $I^s_{\sigma, j}$, but the Hausdorff distances are larger, at $4.46\ (0.59)$ and $2.52\ (0.92)$ respectively. 

Besides the blob detection/GPF method described here, there are similar methods in the transmission electron microscopy community for finding atomic locations that iteratively fit Gaussian peaks with initial means often chosen to be local minima/maxima: for example, Atomap \citep{atomap}, TRACT \citep{levin2020} and mpfit \citep{mpfit2020}. That local minima and their ``intensities'', are used fruitfully in this instance (and stated to have limited value in \citealp{levin2020}) is a testament to the efficacy of the global notion of the size of local minima here, rather than a local one.
 
Comparing these methods with ours yields mean Hausdorff distances of $4.76\ (0.94)$ for the TRACT algorithm and $3.58\ (1.39)$ for Atomap. This is perhaps unsurprising as these algorithms, along the blob detection/GPF approach of \cite{xu2022}, yield subpixel precision for the atomic column position. There is perhaps either not enough noise or sufficient smoothing, so that our algorithm does not shift atomic column positions too drastically from the ground truth, which demonstrates a form of spatial stability of the positive cells of persistent homology. The comparison of outputted intensities of TRACT and blob detection/GPF was done in \cite{xu2022} so we do not replicate it here. 

Blob detection using the Laplacian of Gaussian, used as part of the algorithm in \cite{xu2022} and described in \cite{lindeberg1998} yields images seen in Figures~\ref{f:b25} and S4 
in the Supplementary materials \citep{supp}, after tuning parameters optimally. Even in the case of Image $I_{10, 280}$ (see below for notation and Figure S4), where there are approximately circular blobs present, the method we present here yields results for atomic features that are very similar to the case of blob detection. In conjunction with topological methods for image thresholding such as \emph{PD thresholding} \citep{chung2018}, we may leverage the representation seen in Figure~\ref{f:c25} to binarize our image in a way that accurately preserves shape---see Figure~\ref{f:thr25}. 

PD thresholding was shown to more accurately represent the topology of the underlying image than traditional histogram-based thresholding methods. Here we want to choose the $t$ which maximizes the objective function specified at (11) on p. 1172 of \citeauthor{chung2018}. We choose to only weight $0$-dimensional topological features and ignore $PD_1$. Our results indicate that higher levels of $\sigma$, such as $\sigma=4$, may yield more topologically faithful thresholded images. To visualize these ultra-noisy nanoparticle images more effectively, we will utilize PD thresholding henceforth. We now turn to goodness-of-fit testing of our images and the utilization of topological summaries for extracting signals from videos. 

\begin{figure}
\begin{subfigure}[b]{0.3\textwidth}
\includegraphics[height=1.5in]{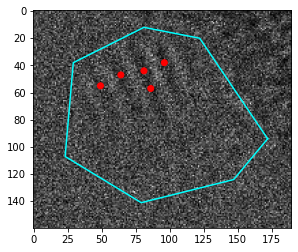}
\caption{}
\label{f:b25}
\end{subfigure}
\begin{subfigure}[b]{0.3\textwidth}
\includegraphics[height=1.5in]{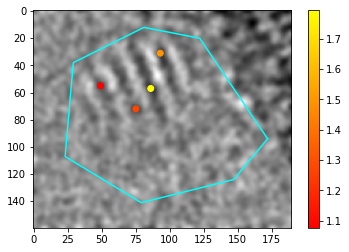}
\caption{}
\label{f:c25}
\end{subfigure}
\begin{subfigure}[b]{0.3\textwidth}
\includegraphics[height=1.5in]{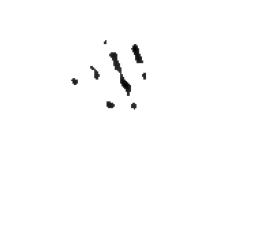}
\caption{}
\label{f:thr25}
\end{subfigure}
\label{f:binarizedIm24}
\caption{In \ref{f:b25} we see the output of LoG blob detection on an image which shows various ridges, or \emph{Miller planes}, in the 2-dimensional nanoparticle image; in \ref{f:c25} we see the same but for our cubical persistence algorithm. In \ref{f:thr25} we see the output of PD thresholding (on the unthresholded image). Though LoG blob detection and our algorithm only outputs points, and Miller planes are visible, there is a one-to-one correspondence between these ridges and points in the case of our algorithm. Thirty values of $\sigma$ from 4 to 8 and a threshold of 0.4 were chosen for blob detection. The values of $\sigma=2$ and $\eta=1$ was chosen for our algorithm in \ref{f:c25}. It did not seem possible to recover 4 planes with blob detection, modifying $\sigma$ and the threshold.}
\end{figure}

\section{Signal detection with hypothesis testing}\label{s:sdht}

\subsection{Setup and assumptions}
We concentrated our analysis on a $N=1124$ frame video of a small area of a catalyst consisting of Pt nanoparticles supported on a larger nanoparticle of $\text{CeO}_2$\footnote{More information on how this data was collected can be found in the Supplementary materials, \citet{supp}.}. As can be seen in Figure \ref{f:b25} and in the left image in Figure~S4 of the Supplementary Materials \citep{supp}, 
the images we aim to analyze are extremely noisy. This necessitates a goodness-of-fit test for a pure noise model. Here we use real-valued summaries of cubical persistence as test statistics for this hypothesis. Let us denote $I_{1}, \dots, I_N$ as the original image sequence with the same image set $\mathcal{I}$. For our basic setup, we consider a series of $m \geq 1$ images summed 
\[
I_{m, \ell}  = \sum_{k=0}^{m-1} I_{\ell+k}, \quad \ell = 1, \dots, N-m+1.
\]
Throughout, let us fix a subimage $\mathcal{L} \subset \mathcal{I}$ and let us suppose that our \emph{unsmoothed} pixels take values on the nonnegative integers. We want to test whether or not the output of the algorithm above produces noise or a definitive signal. We assume that in each image $I_k$, $k = 1, \dots, N$ there is some subset $V_k \subset \mathcal{I}$ that represents the \emph{vacuum}, and as such, is purely constituted of Poisson shot noise, as has been assumed in \cite{levin2020}--- heuristically verified using plotting heuristics for the vacuum region of Pt nanoparticles in \citet{mohan2020}. 

In other imaging contexts we could estimate a null hypothesis of i.i.d noise by sampling from the empirical distribution of pixel values within $R$. However, a Poisson assumption should hold, so we will more rigorously check the Poisson assumption here. If we suppose that $p \in V_{\ell+k}$ for all $k = 1, \dots, m$ so that $I_{\ell+k}(p)$ is Poisson $\mu_{\ell+k}$, then $I_{m, \ell}(p)$ is exactly Poisson with parameter $\sum_{k=0}^{m-1} \mu_{\ell+k}$ if $I_{\ell+k}(p)$ are independent for each $k$. 
At any rate, we assume in our null model that each value $I_{m, \ell}(p)$ has a Poisson distribution for pixels in the vacuum region. It suffices to show that each value $I_{1,\ell}(p)$ has a Poisson distribution, which we will investigate shortly. Throughout this section, identify $R$ with $R \cap \mathcal{L}$---adding any $8$-connected elementary $2$-dimensional cubes lying outside of $R$, so that we may compute cPH. By convention if $d=\infty$, we set the death time $d$ to be the largest pixel value in the rectangular subimage of $I_k$, such as in \citet{chung2018}.

\begin{figure}[t]
\includegraphics[width=3.4in]{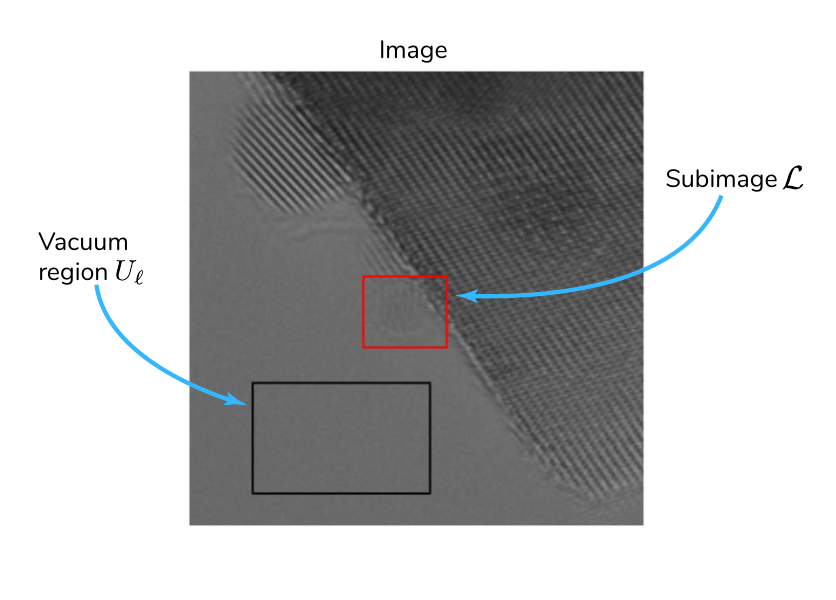}
\caption{Depiction of selection of a subimage and vacuum region for an image $I$. Image depicted here is $I_{m,1}$ with $m=400$. That is, it is the sum of the first 400 frames in a video of \emph{in situ} TEM with 3 Pt nanoparticles. Performing such an averaging is useful to help idenitfy an important subimage which corresponds to interesting atomic structural dynamics.}
\label{f:sub_ill}
\end{figure}

To test whether or not the \emph{observed} output $A(I_{m, \ell})$ coincides with the null hypothesis that the probability distribution $P_\ell \equiv P^m_{\ell}$ which generated the pixel values $\{I_{m,\ell}(p) \big\}_{p \in R}$ in the polygonal region $R$ is equal to the noise distribution $P_{\ell, 0} \equiv P^{m}_{\ell, 0}$, it will help to have some idea of the distribution of $A(I^*_{m, \ell})$ when the random image $I^*_{m, \ell}$ is generated from the noise distribution, i.e. $P_\ell = P_{\ell, 0}$. (Note that $P_\ell$ and $P_{\ell, 0}$ are considered as probability measures on the set $\mathbb{N}_0^R$ of $R$-tuples of nonnegative integers). When the null hypothesis $H_0: P_\ell = P_{\ell, 0}$ holds, for each $p \in R \subset \mathcal{I}$, $I^*_{m, \ell}(p)$ are sampled $\iid$ according to $F_{\ell, 0}$.

As we may not know for certain the entire vacuum region (or the boundary of the vacuum region may change) we may select a subregion $U_\ell \subset V_{\ell+k}$, $k=1,\dots,m$ that we are confident is entirely vacuum for every image $I_{\ell+k}$---see Figure~\ref{f:sub_ill}. For simplicity, we suppose that $U_\ell = U$ for all $\ell = 1, \dots, N$. In the following we identify $U$ with $U \cap \mathbb{Z}^2$. We assume $F_{\ell, 0}$ to be the Poisson distribution with parameter $\lambda_{m,\ell}$. However, there is no evidence that this mean changes (see Figure S1 of the Supplementary Materials, \citealp{supp}), so that $\lambda_{m,\ell} \equiv \lambda_m = m\lambda$ where the maximum likelihood estimate of $\lambda$ is
\[
\hat{\lambda} = \frac{1}{|U| N} \sum_{k=1}^N \sum_{p \in U} I_{k}(p),
\]
assuming independence across frames. In other words, we assume that $P_{\ell, 0} \equiv P_0$ and $F_{\ell, 0} \equiv F_0$. It is worth mentioning that we could have chosen to estimate $\lambda_{m}$ from the mean intensity in $R$ instead, though the intensity would typically be lower, which would indicate a lower variance given the (Poissonian) nature of the data. Therefore, the average lifetimes in the persistence diagrams would be lower and it would reduce the power of the hypothesis tests below. It is also more natural to estimate vacuum behavior from a known vacuum region, rather than treating the data as if \emph{it were} a vacuum region. 

Let us denote $\hat{F}^m_0$ to be the Poisson distribution with parameter $\hat{\lambda}_{m} = m\hat{\lambda}$ and denote $\hat{P}^m_0$ to be the product measure induced by $\hat{F}^m_0$ on $\N_0^R$. 
Denote $F_{\ell, U}$ to be the empirical cdf of the vacuum pixels in frame $\ell$ and $F^m_U$ to be the empirical cdf of 
$$
\big\{ I_{m, \ell}(p): \ell = mk+1, k = 0, \dots, \floor{N/m-1},  p \in U\},
$$
where $F_U \equiv F^1_U$. Let $\hat{F}_0 := \hat{F}^m_0$ be Poisson with mean $\hat{\lambda}$. In practice, we have taken $U$ to be the black rectangular region seen in Figure~\ref{f:sub_ill}, which is 400 pixels by 250 pixels. Here we use the Dvoretzky-Kiefer-Wolfowitz (DKW) inequality \citep[see][]{dasgupta2011, massart1990},
\[
\P( \sup_{k \in \N_0} |F_U(k) - \hat{F}_0(k)| > \epsilon ) \leq 2e^{-2|U|\epsilon^2},
\vspace{-6pt}
\] 
to test the Poisson assumption in the vacuum region. Noting that $\hat{\lambda}$ is a sufficient statistic and that the Kolmogorov-Smirnov (KS) distance is $h_n = 0.00329$, we apply this inequality with $\epsilon = h_n$ and refute the assertion that the data in the vacuum frame are i.i.d. Poisson random variables. 

Though the Poisson assumption may fail to hold precisely over this massive sample, there is little practical and theoretical evidence for doing away with it. Indeed, there are no substantive changes in the Monte Carlo $p$-values when sampling from the empirical distribution $F^{m}_U$ (see Tables~S1 and S2 in the Supplementary materials \citealp{supp}). Furthermore, were we to have a perfect detector, the rate of arrival of electrons at each pixel in the vacuum would follow a Poisson distribution \citep{levin2021direct}. We maintain our initial assumption that the marginal distribution for pixels for the noise distribution in each frame is $\Poi(\lambda)$.
 
 A reasonable hypothesis for what is occurring is the existence of auto- and cross-correlation of pixels between frames, owing to the high temporal resolution. Indeed, the mean autocorrelations over all pixels in the vacuum region $\bar{\rho}(h) = \frac{1}{|U|} \sum_{p \in U} \hat{\rho}_p(h)$ are negative for the first 1093 lags, i.e. $h = 1, \dots, 1093$. If we consider $h \ll N$, under the assumption of independence of each pixel along both spatial\footnote{Empirical semivariograms were checked and we found no evidence to contradict the spatial independence of the pixels---see also Figure~S3 in the Supplementary materials \citep{supp}. 
} and temporal axes, $\hat{\rho}_p(h) \overset{\text{i.i.d.}}{\sim} N(0, 1/1124)$ for all $p \in U$. Therefore, 
\[
\bar{\rho}(h) \simeq N\bigg(0, \frac{1}{1.124\times10^8} \bigg).
\]
Under our hypothesized spatio-temporal independence the probability that \emph{at least one} of the first 50 mean correlations $\bar{\rho}(h)$, $h=1,\dots,50$ is less than its observed value has an upper bound of $2.55\times10^{-10}$, by using the standard probability union bound. Therefore, we can conclude there is strong evidence against temporal independence of the frames in the nanoparticle videos. These negative correlations are small however, with minimum value $-0.001003$ and maximum value $-0.000007$, so summing a small number of frames (such as 10) does not deal a forceful blow to the Poisson assumption of the summed pixels. 

Setting $m=1$, the $p$-values 
\[
\min\bigg\{ 1, 2\exp\Big(-2n\sup_{k \in \N_0} |F_{\ell,U}(k) - \hat{F}_0(k)|^2\Big) \bigg\},
\]
can be shown to be \emph{valid} for each $\ell = 1, \dots, N$ and thus can be used to construct a level $\alpha$ test. Because of the near independence of the frames, we can use the Bonferroni method; there are 2 frames out of 1124 which reject the null hypothesis that the vacuum pixels in frame $\ell$, $\big\{I_{1,\ell}(p)\big\}_{p \in U}$, $\ell = 1, \dots, N$ are i.i.d. $\Poi(\lambda)$ when the significance level $\alpha=0.05$. For practical purposes each frame seems to be identically distributed and very nearly independent, with some distribution that is very close to a Poisson; the KS distances $\sup_{k \in \N_0} |F_{\ell,U}(k) - \hat{F}_{\ell, 0}(k)|$ also have mean nearly $h_n$, and thus satisfy
\[
\frac{1}{N} \sum_{\ell=1}^N \sup_{k \in \N_0} |F_{\ell, U}(k) - \hat{F}_{0}(k)| \approx \sup_{k \in \N_0} |F_U(k) - \hat{F}_0(k)|. 
\]
This is accordant with $\sup_{k \in \N_0} |F_{\ell,U}(k) - \hat{F}_{0}(k)|$ being a stationary and ergodic sequence---suggesting the same for the pixels in the vacuum region of each frame.


Finally, it worth checking how reasonable the independence assumption is for the data in the vacuum. For the images $I_{10, 220}$ and $I_{10, 240}$ above, the empirical semivariogram \cite{cressie1993},  
\[
\hat{\gamma}(l) = \frac{1}{2|N(l)|} \sum_{(p_i, p_j) \in N_l} (I_{m, \ell}(p_i) - I_{m, \ell}(p_j))^2,
\]
with $l = 0, \dots, 14$ and
\[
N(l) := \big\{(p_i, p_j) \in U \times U: \lVert p_i - p_j \rVert \in [4l/3, 4l/3+4/3), \, i \neq j \big\},
\]
indicates a robustly satisfied i.i.d. assumption for the vacuum region in both summed images---see Figure~S3 in the Supplementary materials \citep{supp}. 
At the very least, there is no evidence of correlation. The fact that the plots are nearly indistinguishable lends credibility to the assumption of stationarity across frames as well. Even looking at 
\[
\hat{\gamma}(1) = \frac{1}{2|N(1)|} \sum_{(p_i, p_j): \, \lVert p_i - p_j \rVert =1} (I_{m, \ell}(p_i) - I_{m, \ell}(p_j))^2,
\]
where $N(1)$ is the set we sum over, we see that in the case of $I_{10,220}$ the difference between $\hat{\gamma}(1)$ and the variance is $|4.4983 - 4.4743| = 0.024$ and $|4.45-4.441| = 0.009$ in the case of $I_{10, 240}$. At any rate, it seems as if the parametric approach we have outlined above is tenable, given the theoretical properties of the materials and the physics, even in spite of the fairly minor violations of the assumptions in practice. 

\subsection{Empirical results: hypothesis testing and time series} \label{s:ht_ts}

\begin{figure}
\centering
\begin{subfigure}[b]{0.4\textwidth}
	\centering
	\includegraphics[width=\textwidth]{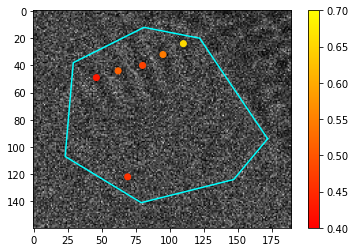}
	\caption{Output $A_{0.4}(I_{10, 240})$ (colored points) superimposed over observed subimage.}
	\label{f:record}
\end{subfigure}
\hfill
\begin{subfigure}[b]{0.4\textwidth}
	\centering
	\includegraphics[width=\textwidth]{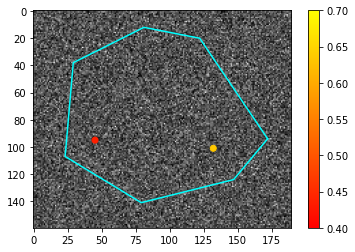} 
	\caption{Output $A_{0.4}(\hat{I}_k)$ (colored points) superimposed over pure noise subimage. \\}
	\label{f:pure_noise}
\end{subfigure}
\caption{The output $A_{0.4}(I)$ for an observed subimage (left) and a pure noise/simulated subimage (right). Here $I_{10, 240}$ can be seen to have a strong signal corresponding to various atomic features.}
\label{f:output_cyan}
\end{figure}

Assume we believe that the number of (thresholded) columns output by our algorithm for the summed image $I_{m, \ell}$ is higher for an image with a strong signal in contrast to an image that is composed entirely of noise. In other words, we are interested in the quantity $n_\ell := |A_\eta(I_{m, \ell})|$, and in particular the evidence of $n_\ell$ against $H_0$. We follow \citet{davison_hinkley} in constructing a $p$-value for our Monte Carlo hypothesis test. First, we generate pixel values in $\mathcal{L}$ (hence $R$ as well) according to $\hat{P}^m_0$ to yield an image $\hat{I}^*_{m, \ell}$. We proceed by generating a number of i.i.d. instances of $\hat{I}^*_{m, \ell}$ denoted $\hat{I}_1, \dots, \hat{I}_n$. There is precedent to the idea of using generated simulated smoothed images (which can be considered as discrete random fields) for hypothesis testing---an example of which can be seen in \cite{taylor2007maxima}. The initial $p$-value we aim to estimate is
\begin{equation} \label{e:pval}
\P_0 \big( |A_\eta(I^*_{m, \ell})| \geq n_\ell \big), 
\end{equation}
where $\P_0$ is a probability measure under which the null hypothesis holds. The $p$-value \eqref{e:pval} is valid when we condition on $\hat{\lambda}$. In practice, we estimate the true $p$-value \eqref{e:pval} by the rank of $n_\ell$ amongst the simulated value $|A(\hat{I}_k)|$. As $n_\ell$ is discrete (integer-valued), such a rank is not unique, so we utilize the Monte Carlo $p$-value 
\[
\frac{1}{n+1} \bigg(1 + \sum_{i=1}^n \ind{ |A_\eta(\hat{I}_k)| \geq n_\ell } \bigg).
\]
This is reasonable if both $|U_\ell|$ and $n$ are sufficiently large, by \eqref{e:pval}. For our particular setting, let $m = 10$, $\sigma=2,4,6$, $\eta = t(\sigma)$ and set $n = 9999$, $U$ to be the black region in Figure~\ref{f:sub_ill}, and $R$ to be the cyan polygon in Figure~\ref{f:output_cyan}---where $t(\sigma)$ is a function calibrated by the user to recover relevant topological features in an image. Here we have $t(2) = 1$, $t(4) = 0.4$, and $t(6) = 0.1$. For $\sigma = 4$ and $\ell=240$, we estimate that 
\[
\P_0 \big( |A_\eta(I^*_{m, \ell})| \geq n_\ell \big) = 0.0001
\]
which yields very strong evidence against $H_0$ for the pixels in $R$ in Figure~\ref{f:record}, where $n_\ell = 6$. This method worked well because there are many ``significant'' features in the nanoparticle image and would also work with any algorithm $A$ which outputs a (marked) point process, such as blob detection \citep{lindeberg1998, kong_blob}. However, if there is only one or two highly persistent features, this test will be decidedly underpowered. 
Additionally, there is issue of a choosing a threshold $\eta$---there are many other values of $\eta$ that would lead to the anticipated rejection of $H_0$; also, manually inspecting thousands of images to find relevant features is infeasible in practice. In general, we may consider a real-valued functional $f$ of a marked point process on $R$, such that larger values of $f\big(A(I_{m, \ell})\big)$ for our summed image would lead us to reject $H_0$. We can proceed with the exact same framework as the above, but what sort of function would yield useful information? In principle, we could use any method that can be used with a point process, see \citet{illian2008}. But such methods could be used with the output of methods such as blob detection \citep[as in][]{xu2022}, as well. As such, let us consider a real-valued function of the marks of $A(I)$---i.e. the lifetimes of the points in the persistence diagram $PD_0$---called \emph{persistent entropy} \citep{rucco2016, stab_pers_entr}. Because we consider output of algorithm in the case of noise to be associated with more disorder, we actually take the negative of the persistence entropy, or 
\[
H\big(A(I)\big) := \sum_{(p, l) \in A(I)} l/L \log( l/L )= \sum_{(b,d) \in PD_0} \big((d-b)/L\big) \log\big((d-b)/L\big) ,
\]
where $L = \sum_{(p,l) \in A(I)} l = \sum_{(b,d) \in PD_0} d-b$. Higher values of $H\big(A(I)\big)$, i.e. values closer to zero, signify smaller entropy. Using the negative of persistent entropy yields a Monte Carlo $p$-value of 
\[
\P_0 \big( H\big(A(I^*_{m, \ell})\big) \geq H_\ell \big) = 0.0317,
\]
where $H_\ell = H\big(A(I_{m, \ell})\big)$ is again the observed value for image depicted in Figure~\ref{f:record}. One may also consider the longest barcode (or, the infinity norm), i.e.
\[
L\big(A(I)\big) := \max \{l: (p, l) \in A(I)\} = \max \{d-b: (b,d) \in PD_0\}
\]
as well as the sample mean persistence $E(A(I)) =  \sum_{(p, l) \in A(I)} l/ |A(I)|$. We also introduce the ALPS (accumulated lifetimes of persistence survival) statistic, $\Delta(A(I))$, defined by 
\begin{equation}\label{e:alps}
\Delta(A(I)) := \int_0^{\infty} \log U(\eta) \dif{\eta},
\end{equation}
where $U(\eta) = \sum_{(p,l) \in A(I)} \ind{l > \eta} = \sum_{(b,d) \in PD_0} \ind{d-b > \eta}$. We can easy generalize this to any persistence diagram $PD_k$, just as one can with any of the aforementioned summaries. Order the lifetimes of $PD_0$ as $l_{(1)} \leq \cdots \leq l_{(K)}$ where $K = |A(I)|$ is the number of points/barcodes in the persistence diagram. There is another convenient representation of \eqref{e:alps}, which is worth mentioning.
\begin{proposition}\label{p:alps}
\[
\Delta(A(I)) = - \sum_{i=1}^{K-1} l_{(i)} \log\bigg( 1- \frac{1}{K-i+1} \bigg),
\]
\end{proposition}

We offer a short proof in the Supplementary materials \citep{supp}. Based on Proposition~\ref{p:alps} there is no need to consider how to treat the infinite barcode, if it is present in $R$. The ALPS statistic is similar in spirit to the accumulated persistence function, and aims to balance the information content of the longest barcode with that of the lifetime sum, or 1-norm, $\sum_{i=1}^K l_{(i)}$ of the persistence diagram. It also bears more than a passing resemblance to persistent entropy, which could explain why they act so well as topological summaries. 

Other summaries contain useful information but fluctuate wildly\footnote{Tables S3--S6, supporting this conclusion, can be seen in the Supplementary materials \citealp{supp}.}, such as the sample skewness of the persistence lifetimes, or never yield a significant signal---e.g. the $p$-norms of persistence \citep{cohensteiner2010} and the signal-to-noise ratio (of mean lifetime divided by standard deviation of lifetimes). The persistent entropy was found to be most stable to whether or not we were in parametric or nonparametric setting. The ALPS statistic $p$-values decreased across the board in the nonparametric setting, which is what one would expect as in both images $I_{10, 220}$ and $I_{10, 240}$ there appears to be some signal. The summary of the Monte Carlo $p$-values for each of these five test statistics can be seen in Table~\ref{t:mcpv} and \ref{t:mcpv2}.


\begin{table}[h]
\resizebox{\columnwidth}{!}{
\bgroup
\def\arraystretch{1.2}
\begin{tabular}{|l|r|r|r|r|r|}
\hline
\multirow{2}{*}{Test statistic} & Number of columns & Persistent entropy & Longest barcode & Mean persistence & ALPS statistic \\
& $|A_{t(\sigma)}(I_{10, 240})|$ & $H\big(A(I_{10, 240})\big)$ & $L\big(A(I_{10, 240})\big)$ & $E\big(A(I_{10, 240})\big)$ & $\Delta\big(A(I_{10, 240})\big)$ \\ 
\hline
$\sigma = 2$ & 0.0001 & 0.0001 & 0.0595 & 0.2303 & 0.0008 \\
\hline
$\sigma = 4$ & 0.0001 & 0.0317 & 0.0063 & 0.0013 & 0.0001\\
\hline
$\sigma = 6$ & 0.1140 & 0.7966 & 0.5573 & 0.2660 & 0.1032 \\
\hline
\end{tabular}
\egroup
}
\vspace{12pt}
\caption{Monte Carlo $p$-values $p_n$ for various real-valued topological summaries of $I_{10, 240}, \hat{I}_1, \dots, \hat{I}_{9999}$ smoothed with varying values of $\sigma$. Images were generated according to $\hat{P}_{\ell, 0}$. With at least 95\% confidence the true $p$-value lies in $p_n \pm 0.0137$, truncating at 0 or 1---see Supplementary materials \citep{supp} for more information.}
\label{t:mcpv}
\end{table}

\begin{figure}[h]
\centering
\begin{subfigure}[b]{0.4\textwidth}
	\centering
	\includegraphics[width=\textwidth]{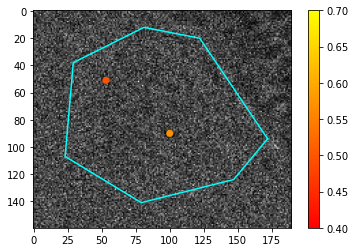}
	\caption{Output $A_{0.4}(I_{10, 220})$ (colored points) superimposed over observed subimage.}
	\label{f:record2}
\end{subfigure}
\hfill
\begin{subfigure}[b]{0.4\textwidth}
	\centering
	\includegraphics[width=\textwidth]{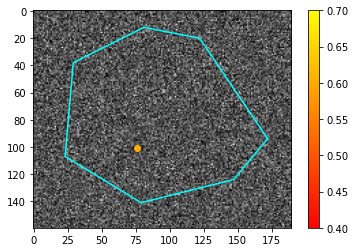} 
	\caption{Output $A_{0.4}(\hat{I}_k)$ (colored points) superimposed over pure noise subimage. \\}
	\label{f:pure_noise2}
\end{subfigure}
\caption{The output $A_{0.4}(I)$ for an observed subimage (left) and a pure noise/simulated subimage (right). Here $I_{10, 220}$ can be seen to have a weak signal corresponding to various atomic features.}
\label{f:output_cyan2}
\end{figure}

In the image in Figure~\ref{f:record}, as there is an extremely robust nanoparticle structure present, we would expect to reject the null hypothesis of an the image consisting of i.i.d. Poisson random variables. 
If we examine a different summed image, as in Figure~\ref{f:record2}, we can see upon thresholding the image with $\eta = 0.4$ that there are only two columns. Hence the larger estimated $p$-value in this case. However, both persistence entropy $H$ and longest barcode $L$ indicate that it is reasonable to reject the null hypothesis that $P_\ell = P_0$---see Table~\ref{t:mcpv2}.

\begin{table}[h]
\resizebox{\columnwidth}{!}{
\bgroup
\def\arraystretch{1.2}
\begin{tabular}{|l|r|r|r|r|r|}
\hline
\multirow{2}{*}{Test statistic} & Number of columns & Persistent entropy & Longest barcode & Mean persistence & ALPS statistic \\
& $|A_{t(\sigma)}(I_{10, 220})|$ & $H\big(A(I_{10, 220})\big)$ & $L\big(A(I_{10, 220})\big)$ & $E\big(A(I_{10, 220})\big)$ & $\Delta\big(A(I_{10, 220})\big)$ \\
\hline
$\sigma = 2$ & 0.0916 & 0.0604 & 0.1485 & 0.3523 & 0.0851 \\
\hline
$\sigma = 4$ & 0.2015 & 0.0213 & 0.0378 & 0.0051 & 0.0531 \\
\hline
$\sigma = 6$ & 0.2888 & 0.2040 & 0.1620 & 0.0328 & 0.1762 \\
\hline
\end{tabular}
\egroup
}
\vspace{12pt}
\caption{Monte Carlo $p$-values $p_n$ for various real-valued topological summaries of $I_{10, 220}, \hat{I}_1, \dots, \hat{I}_{9999}$ smoothed with varying values of $\sigma$. Images were generated according to $\hat{P}_{\ell, 0}$. With at least 95\% confidence the true $p$-value lies in $p_n \pm 0.0137$ (truncating at 0 or 1)---see Supplementary materials \citep{supp} for more information.}
\label{t:mcpv2}
\end{table}

\begin{figure}[h]
	\centering
	\includegraphics[width=2.5in]{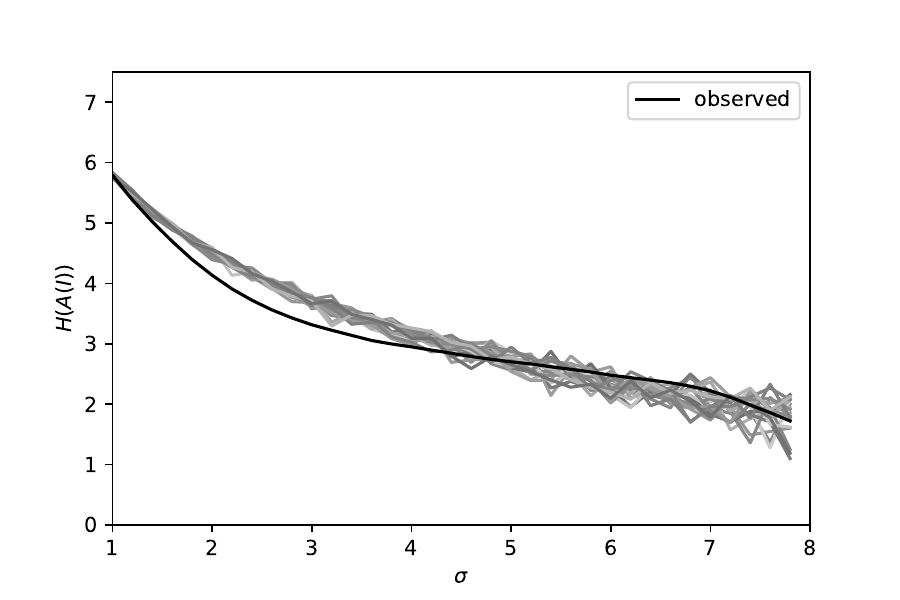}
	\includegraphics[width=2.5in]{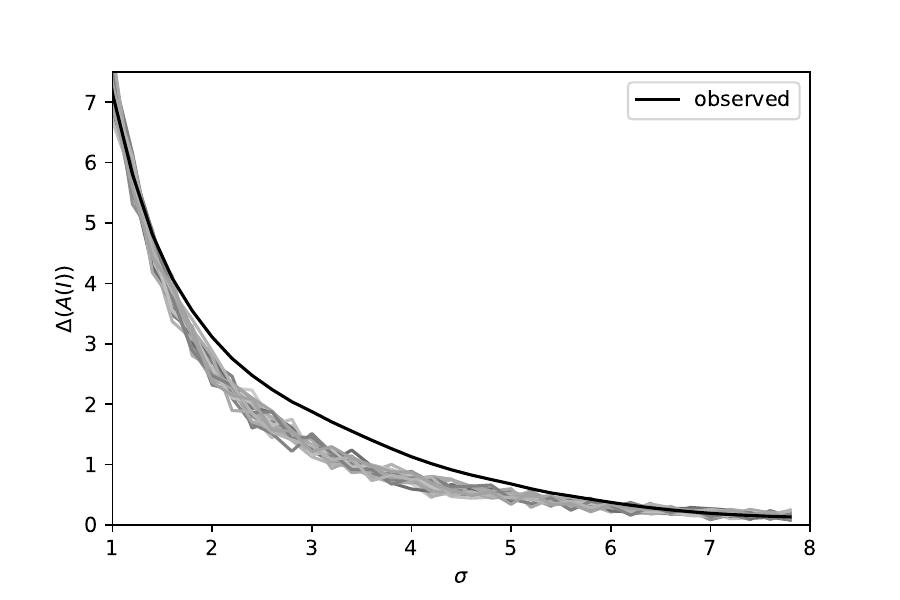}
	\caption{Persistent entropy (left) and ALPS (right) for various smoothing parameters $\sigma$ for image $I_{10, 240}$, compared with same series for 19 simulated images. }
	\label{f:pers_entr24} 
\end{figure}

\begin{figure}[t]
\includegraphics[width=5.5in]{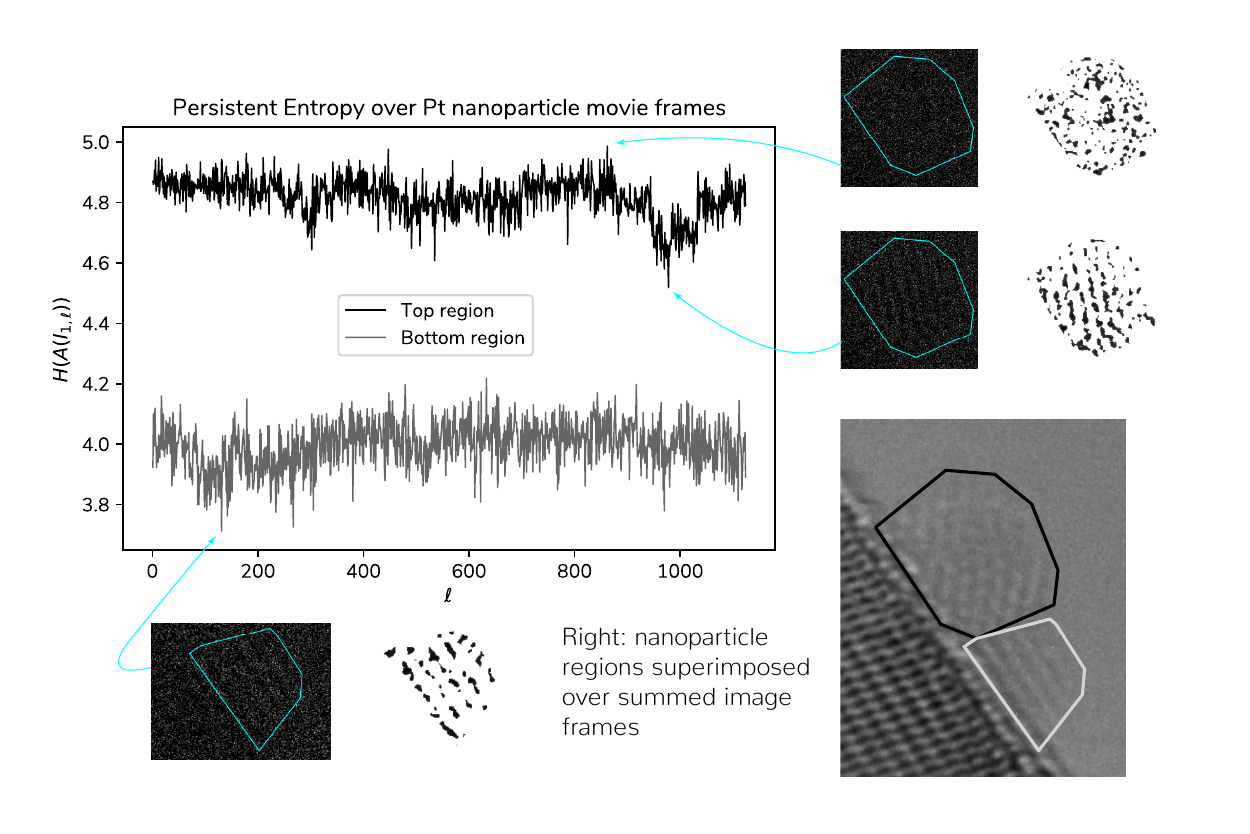}
\caption{Time series of persistent entropy over all 1124 frames for the nanoparticle regions depicted in lower right. Selected nanoparticle frames corresponding to high/low values of persistent entropy are depicted along with binarized images derived from PD thresholding.}
\label{f:nano_region}
\end{figure}

Based on the output of the tables above, it seems that the ALPS statistic and persistent entropy are the best metrics of the five that we've considered. The ALPS statistic has the smallest variance across the $p$-values in the tables in the above; it additionally seems to correlate fairly strongly with the $p$-value associated to the thresholded number of columns. Furthermore, it does not require any tuning of the threshold parameter $\eta$. Both the ALPS statistic and the persistent entropy seems to yield the best ``separation'' between an observed image with clear signal and a noisy simulated image---see Figure~\ref{f:pers_entr24}. Persistent entropy also corresponds with known changes in atomic configurations (see Figure~\ref{f:nano_region}) and enjoys various stability properties \citep{stab_pers_entr}. Hence, for smoothing parameters $\sigma$ in any compact interval bounded away from zero, it can be shown that $\sigma \mapsto I_{\sigma}$ is a smooth function as well, thus persistent entropy does not fluctuate too rapidly for small changes in $\sigma$ (one can use the stability theorem for cubical persistence from \citealp{skraba2021wasserstein}). Experimentally, it appears this holds with the ALPS statistic too---see Figure~\ref{f:pers_entr24}. 
 
\begin{figure}[t]
	\includegraphics[width=3.5in]{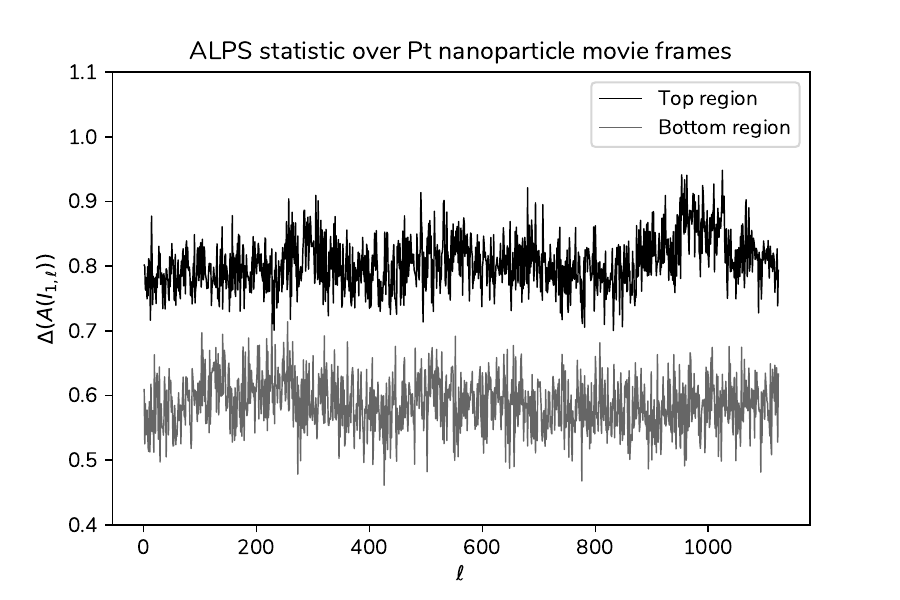}
	\caption{Time series of ALPS statistic over all 1124 frames for nanoparticle regions depicted in lower right of Figure~\ref{f:nano_region}.}
	\label{f:alps_series}
\end{figure} 

\begin{figure}[h!]
    \centering
    \includegraphics[width=3.5in]{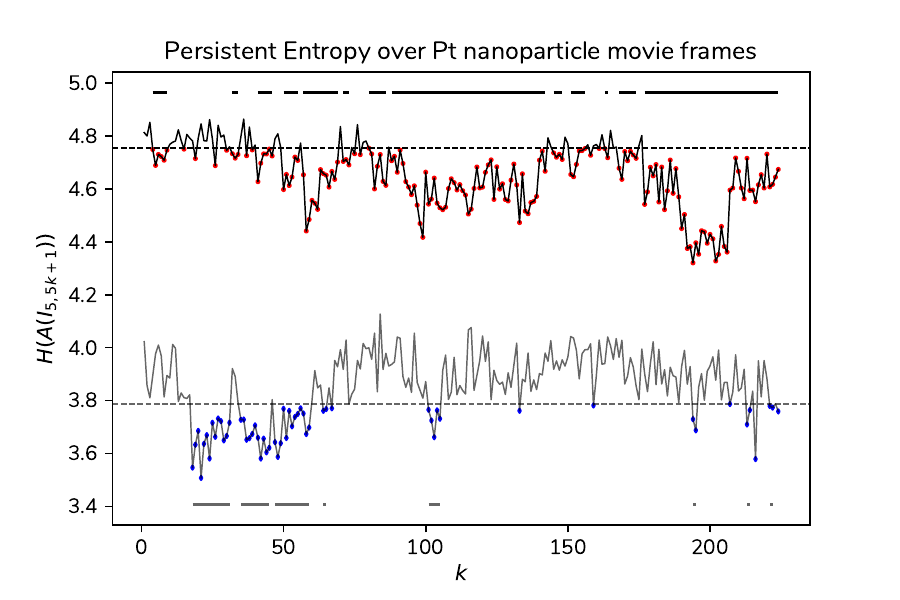}
    \caption{Persistent entropy for $I_{5, 5k+1}$, $k=0, 1, \dots, 223$ image sequence for top and bottom regions with non-noisy frames marked in red and blue at the $\alpha=0.05$ level. Runs of significant frames for each region depicted above and below the plotted curves.}
    \label{f:pe_bh}
\end{figure}

\begin{figure}[t]
    \centering
    \includegraphics[width=3.5in]{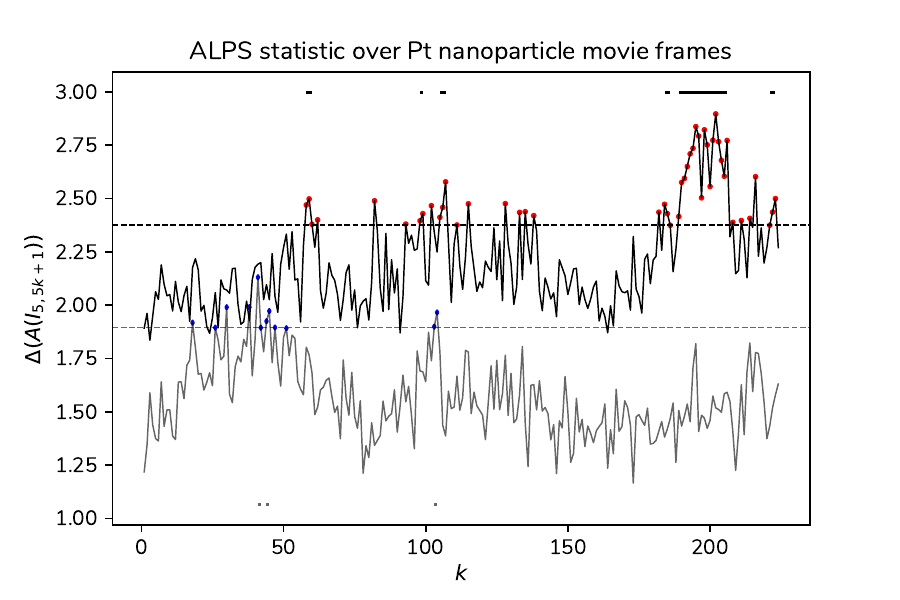}
    \caption{ALPS statistic for $I_{5, 5k+1}$, $k=0, 1, \dots, 223$ image sequence for top and bottom regions with non-noisy frames marked in red and blue at the $\alpha=0.05$ level. Runs of significant frames for each region depicted above and below the plotted curves.}
    \label{f:alps_bh}
\end{figure}


In Figures~\ref{f:nano_region} and \ref{f:alps_series}, we derive time series by calculating the topological summaries persistent entropy and ALPS statistic, respectively, for the two nanoparticle regions seen in the lower right portion of Figure~\ref{f:nano_region}). The summaries were calculated framewise and $\sigma=2$ was chosen as it led to a high degree of correspondence with ``stable'' states, meaning presence of visible nanoparticle structure. Such phenomena began to disappear as $\sigma$ increased, and the association of the statistic between low values (persistent entropy) or high values (ALPS) and high order reversed when $\sigma$ was too large. Part of the reason for this is that noisy images with a sufficiently large degree of smoothing will have fewer total outputted points in $A(I)$, which typically leads to less variation and entropy. An illustration of this can be seen in Figure~\ref{f:pers_entr24}. 

\subsection{Multiple testing using persistent entropy \\ and the ALPS statistic}

Using the extension of the Benjamini-Hochberg procedure \citep{bh, benj_yeku}, we test each of the frames in the summed series $I_{5, 5k+1}$, $k = 0, 1, \dots, 223$ versus the null hypothesis that they were generated according to $P_{\ell, 0}$, with $m=5$. We use $m=5$ here instead of the individual frames, because very few frames (only for the top region, using persistent entropy) are significant. We proceed by generating $n=9999$ images according to the same recipe outlined at the start of Section~\ref{s:ht_ts}. Again we make the assumption that $P_{\ell, 0} = P_0$, or that the same noise hypothesis is shared by each frame. This is not unreasonable based on what we have found so far. Our approach is at least somewhat similar to the framework described for multiple testing using persistent homology described in \citet{vejdemo2020}. However, in said article, they are concerned with point cloud-based rather than cubical homology and potentially different null distributions for each test. Here our hypotheses are $H_0: P_{5k+1} = P_0$, for each $k=0,1, \dots, 223$. 

As we generate so few images owing to the stationarity assumption of our time series of topological summaries, the computational costs are signficantly less than that of other hypothesis testing settings using persistent homology \citep{vejdemo2020, robinson2017}. We also guarantee that our multiple testing framework has asymptotic false discovery rate less than $\alpha$, almost surely---see Proposition~\ref{p:fdrn}. 

Figures~\ref{f:pe_bh} and \ref{f:alps_bh} indicate that the persistent entropy has much greater power than the ALPS statistic, though depending on the application it may be a bit overzealous. That is to say, for researchers that are more conservative in their desire to detect atomic features, the ALPS statistic may be preferable. It is also possible to normalize the time series and consider the convex combination $\lambda H(A(I)) + (1-\lambda)\Delta(A(I))$ for $\lambda \in (0,1)$. One may verify that for the values of $k$ immediately preceding 200 for the top region and values of $k$ around 25 for the bottom region, there is a significant nanoparticle structure present---see the Supplementary materials \citep{supp}, Section S5, and Figures S5b and S5c.

\section{Discussion}

In this paper we have discussed a novel means of detecting atomic features in nanoparticle images, which compares favorably to existing methods and is nonparametric in its estimation of intensity. We have also investigated means of hypothesis testing the presence of signals in images with ultra-low signal-to-noise ratio and detailed a useful method for deriving time series from noisy videos, which yields a new topological change-point detection method. We have also introduced the ALPS statistic, which conveys much of the same information content in a hypothesis testing context as the number of columns present after appropriate thresholding. 

As topological data analysis is such a young field, there is no shortage of directions along which the methods in this article could be expanded. For example, one could employ a functional version of $|A_\eta(I)|$; alternatively, some another functional summary of 
persistence diagrams may hold promise, especially used in conjunction with a global rank envelope test as in \citet{biscio_acf} and \citet{myllymaki2017}. Furthermore, using persistent homology for point clouds such as the Vietoris-Rips filtration \citep{geom_top2018}, would utilize the location information derived from the above algorithm in an essential way. Using a weighted Vietoris-Rips filtration may even furnish more precise results \citep{anai2020dtm}, as the marked point process output of the algorithm naturally yields weights for each point. After deriving an additional persistence diagram from the output $A(I)$, one may choose an appropriate functional summary \citep[see][]{berry2020} and proceed from there.

Furthermore, other choices of filtrations that are more suited to capturing geometric information, may be considered in future studies. Examples include treating the locations of black pixels as embedded into Euclidean space, and then applying a Vietoris-Rips or density-based filtration \citep{garin2019, turkes2021}. As mentioned earlier, these methods require the choice of a threshold and hence multi-parameter persistent homology may be a more appropriate tool---see \cite{chung2022}. In sum, our hope is that this study can find adoption in the microscopy community and also be used as point of departure for future studies at the intersection of image processing, statistics, and TDA.

\section{Appendix}
Fix a probability space $(\Omega, \mathcal{F}, \P)$. We consider the problem of multiple testing in a Monte Carlo setting where $H_{0,1}, \dots, H_{0,N}$ are null hypotheses under which there exist distribution functions $(F_{0,k})_{k=1}^N$ satisfying $F_{0, k} = F_0$ for all $k$, for $F_0$ some continuous cdf on $\R$. Here $N$ is considered fixed but arbitrary. Denote our (i.i.d.) Monte Carlo sample of test statistics under $F_0$ as $T_1, \dots, T_n$, which we assume are also independent from the observed (random) test statistics $t_1, \dots, t_N$. Let $p_{n, 1}, \dots, p_{n,N}$ denote our Monte Carlo $p$-values and $p_{n, (1)} \leq \cdots \leq  p_{n, (N)}$ to be their order statistics. We specify $H_{0, (1)}, \dots, H_{0, (N)}$ to be the hypotheses corresponding to the ordered $p$-values. We suppose that all of our test statistics are nonnegative. Let us define
\[
p_{n, k} := \frac{1}{n+1}\bigg( 1 + \sum_{i=1}^n \ind{T_i \geq t_k}\bigg),
\]
and $r_{n,k}$ to be the rank of $p_{n,k}$ among all of the values of $p_{n,1}, \dots, p_{n,N}$, e.g. $r_{n,(k)} = k$. We construct a test (a Monte Carlo version of the one in \citealp{benj_yeku}), based off of the inequalities 
\begin{equation}\label{e:hyp_test}
p_{n,k} \leq \frac{r_{n,k} \alpha}{N C_N},
\end{equation}
where $C_N  = \sum_{k=1}^N 1/k$. Let $\text{FDR}_n$ be the \emph{false discovery rate} for the test based on \eqref{e:hyp_test}. That is, $\text{FDR}_n$ is the expected proportion of falsely rejected hypotheses over the total number of rejected hypotheses due to the criterion in \eqref{e:hyp_test} (equal to 0 if no hypotheses are rejected). Note that $\text{FDR}_n$ is conditional on $(T_i)_{i=1}^n$. Let $\text{FDR}$ denote the false discovery rate based on the true $p$-values $p_k = \P(T_i \geq t_k)$. We specify the exact nature of test---along with a theoretical guarantee---in the next proposition.

\begin{proposition}\label{p:fdrn}
Suppose $T_1, \dots, T_n$ are test statistics sampled i.i.d. according to continuous cdf $F_0$. If the $(t_k)_{k=1}^N$ have continuous distributions under the alternative hypotheses as well, then the test which rejects $H_{0, (1)}, \dots, H_{0, (\ell)}$ if 
\[
\ell = \sup \big\{k = 1, \dots, N: p_{n, (k)} \leq k\alpha/(N C_N)\big\}
\]
satisfies $\mathrm{FDR}_n \overset{\mathrm{a.s.}}{\to} \mathrm{FDR}, \ n \to \infty$. Thus,
\[
\limsup_n \mathrm{FDR}_n \leq \alpha, \quad \mathrm{a.s.}
\]
\end{proposition}

The proof of Proposition~\ref{p:fdrn} can be seen in the Supplementary materials \citep{supp}. We are also able to consider the test statistics $t_i$, $i=1,\dots, N$ as random as well. Our result is an improvement on Corollary 1 in \cite{gandy_hahn} for the Benjamini-Hochberg procedure, in that our results holds with probability 1, rather than probability $1-\epsilon$, for $n$ large enough.

\bibliography{AtomSurfTDA}

\begin{thebibliography}{}

\bibitem[Anai et~al., 2020]{anai2020dtm}
Anai, H., Chazal, F., Glisse, M., Ike, Y., Inakoshi, H., Tinarrage, R., and
  Umeda, Y. (2020).
\newblock {DTM-based filtrations}.
\newblock In {\em Topological Data Analysis}, pages 33--66. Springer.

\bibitem[Atienza et~al., 2020]{stab_pers_entr}
Atienza, N., Gonz{\'a}lez-D{\'\i}az, R., and Soriano-Trigueros, M. (2020).
\newblock On the stability of persistent entropy and new summary functions for
  topological data analysis.
\newblock {\em Pattern Recognition}, 107:107509.

\bibitem[Babaud et~al., 1986]{babaud1986}
Babaud, J., Witkin, A.~P., Baudin, M., and Duda, R.~O. (1986).
\newblock {Uniqueness of the Gaussian kernel for scale-space filtering}.
\newblock {\em IEEE transactions on pattern analysis and machine intelligence},
  (1):26--33.

\bibitem[Benjamini and Hochberg, 1995]{bh}
Benjamini, Y. and Hochberg, Y. (1995).
\newblock Controlling the false discovery rate: a practical and powerful
  approach to multiple testing.
\newblock {\em Journal of the Royal statistical society: series B
  (Methodological)}, 57(1):289--300.

\bibitem[Benjamini and Yekutieli, 2001]{benj_yeku}
Benjamini, Y. and Yekutieli, D. (2001).
\newblock {The control of the false discovery rate in multiple testing under
  dependency}.
\newblock {\em The Annals of Statistics}, 29(4):1165 -- 1188.

\bibitem[Berry et~al., 2020]{berry2020}
Berry, E., Chen, Y.-C., Cisewski-Kehe, J., and Fasy, B.~T. (2020).
\newblock {Functional summaries of persistence diagrams}.
\newblock {\em Journal of Applied and Computational Topology}, 4(2):211--262.

\bibitem[Biscio and M{\o}ller, 2019]{biscio_acf}
Biscio, C.~A. and M{\o}ller, J. (2019).
\newblock The accumulated persistence function, a new useful functional summary
  statistic for topological data analysis, with a view to brain artery trees
  and spatial point process applications.
\newblock {\em Journal of Computational and Graphical Statistics},
  28(3):671--681.

\bibitem[Blumberg et~al., 2014]{blumberg2014}
Blumberg, A.~J., Gal, I., Mandell, M.~A., and Pancia, M. (2014).
\newblock Robust statistics, hypothesis testing, and confidence intervals for
  persistent homology on metric measure spaces.
\newblock {\em Foundations of Computational Mathematics}, 14(4):745--789.

\bibitem[Boissonnat et~al., 2018]{geom_top2018}
Boissonnat, J.-D., Chazal, F., and Yvinec, M. (2018).
\newblock {\em Geometric and topological inference}, volume~57.
\newblock Cambridge University Press.

\bibitem[Cericola et~al., 2017]{cericola2017}
Cericola, C., Johnson, I.~J., Kiers, J., Krock, M., Purdy, J., and Torrence, J.
  (2017).
\newblock Extending hypothesis testing with persistent homology to three or
  more groups.
\newblock {\em Involve, a Journal of Mathematics}, 11(1):27--51.

\bibitem[Chen and Edelsbrunner, 2011]{chen2011diffusion}
Chen, C. and Edelsbrunner, H. (2011).
\newblock Diffusion runs low on persistence fast.
\newblock In {\em 2011 International Conference on Computer Vision}, pages
  423--430. IEEE.

\bibitem[Chen et~al., 2020]{chen2020}
Chen, X., Chen, D., Weng, M., Jiang, Y., Wei, G.~W., and Pan, F. (2020).
\newblock {Topology-Based Machine Learning Strategy for Cluster Structure
  Prediction}.
\newblock {\em Journal of Physical Chemistry Letters}, 11(11):4392--4401.

\bibitem[Chintakunta et~al., 2015]{pers_entropy}
Chintakunta, H., Gentimis, T., Gonzalez-Diaz, R., Jimenez, M.-J., and Krim, H.
  (2015).
\newblock An entropy-based persistence barcode.
\newblock {\em Pattern Recognition}, 48(2):391--401.

\bibitem[Chung and Day, 2018]{chung2018}
Chung, Y.-M. and Day, S. (2018).
\newblock Topological fidelity and image thresholding: A persistent homology
  approach.
\newblock {\em Journal of Mathematical Imaging and Vision}, 60(7):1167--1179.

\bibitem[Chung et~al., 2022]{chung2022}
Chung, Y.-M., Day, S., and Hu, C.-S. (2022).
\newblock A multi-parameter persistence framework for mathematical morphology.
\newblock {\em Scientific Reports}, 12(1):6427.

\bibitem[Cohen-Steiner et~al., 2007]{cohenstein_stab}
Cohen-Steiner, D., Edelsbrunner, H., and Harer, J. (2007).
\newblock {Stability of persistence diagrams}.
\newblock {\em Discrete and Computational Geometry}, 37(1):103--120.

\bibitem[Cohen-Steiner et~al., 2010]{cohensteiner2010}
Cohen-Steiner, D., Edelsbrunner, H., Harer, J., and Mileyko, Y. (2010).
\newblock Lipschitz functions have $l_p$-stable persistence.
\newblock {\em Foundations of computational mathematics}, 10(2):127--139.

\bibitem[Cressie, 1993]{cressie1993}
Cressie, N. (1993).
\newblock {\em Statistics for spatial data}.
\newblock John Wiley \& Sons.

\bibitem[DasGupta, 2011]{dasgupta2011}
DasGupta, A. (2011).
\newblock {\em Probability for statistics and machine learning: fundamentals
  and advanced topics}.
\newblock Springer Science \& Business Media.

\bibitem[Davison and Hinkley, 1997]{davison_hinkley}
Davison, A.~C. and Hinkley, D.~V. (1997).
\newblock {\em Bootstrap Methods and their Application}.
\newblock Cambridge Series in Statistical and Probabilistic Mathematics.
  Cambridge University Press.

\bibitem[D\l{}otko, 2015]{gudhi_cubical}
D\l{}otko, P. (2015).
\newblock Cubical complex.
\newblock In {\em {GUDHI} User and Reference Manual.} {GUDHI Editorial Board}.

\bibitem[Edelsbrunner et~al., 2002]{edelsbrunner2002}
Edelsbrunner, Letscher, and Zomorodian (2002).
\newblock {Topological Persistence and Simplification}.
\newblock {\em Discrete {\&} Computational Geometry}, 28(4):511--533.

\bibitem[Edelsbrunner and Harer, 2010]{edelsbrunner2010}
Edelsbrunner, H. and Harer, J. (2010).
\newblock {\em {Computational topology: an introduction}}.
\newblock American Mathematical Society, Providence, Rhode Island.

\bibitem[Egerton, 2013]{egerton2013}
Egerton, R. (2013).
\newblock Control of radiation damage in the {TEM}.
\newblock {\em Ultramicroscopy}, 127:100--108.

\bibitem[Egerton, 2019]{egerton2019}
Egerton, R. (2019).
\newblock Radiation damage to organic and inorganic specimens in the {TEM}.
\newblock {\em Micron}, 119:72--87.

\bibitem[Egerton et~al., 2004]{egerton2004}
Egerton, R., Li, P., and Malac, M. (2004).
\newblock Radiation damage in the {TEM} and {SEM}.
\newblock {\em Micron}, 35(6):399--409.

\bibitem[Faruqi and McMullan, 2018]{faruqi2018}
Faruqi, A. and McMullan, G. (2018).
\newblock Direct imaging detectors for electron microscopy.
\newblock {\em Nuclear Instruments and Methods in Physics Research Section A:
  Accelerators, Spectrometers, Detectors and Associated Equipment},
  878:180--190.

\bibitem[Fasy et~al., 2014]{fasy2014}
Fasy, B.~T., Lecci, F., Rinaldo, A., Wasserman, L., Balakrishnan, S., and
  Singh, A. (2014).
\newblock {Confidence sets for persistence diagrams}.
\newblock {\em The Annals of Statistics}, 42(6):2301--2339.

\bibitem[Gandy and Hahn, 2014]{gandy_hahn}
Gandy, A. and Hahn, G. (2014).
\newblock Mmctest—a safe algorithm for implementing multiple monte carlo
  tests.
\newblock {\em Scandinavian Journal of Statistics}, 41(4):1083--1101.

\bibitem[Garin et~al., 2020]{garin2020duality}
Garin, A., Heiss, T., Maggs, K., Bleile, B., and Robins, V. (2020).
\newblock {Duality in Persistent Homology of Images}.
\newblock {\em arXiv preprint arXiv:2005.04597}.

\bibitem[Garin and Tauzin, 2019]{garin2019}
Garin, A. and Tauzin, G. (2019).
\newblock {A topological 'reading' lesson: Classification of MNIST using TDA}.
\newblock {\em Proceedings - 18th IEEE International Conference on Machine
  Learning and Applications, ICMLA 2019}, pages 1551--1556.

\bibitem[Getreuer, 2013]{getreuer2013}
Getreuer, P. (2013).
\newblock A survey of gaussian convolution algorithms.
\newblock {\em Image Processing On Line}, 2013:286--310.

\bibitem[Gillies, 2013]{shapely}
Gillies, S. (2013).
\newblock The shapely user manual.
\newblock {\em URL https://pypi. org/project/Shapely}.

\bibitem[Illian et~al., 2008]{illian2008}
Illian, J., Penttinen, A., Stoyan, H., and Stoyan, D. (2008).
\newblock {\em Statistical analysis and modelling of spatial point patterns},
  volume~70.
\newblock John Wiley \& Sons.

\bibitem[Jiang et~al., 2021]{jiang2021}
Jiang, Y., Chen, D., Chen, X., Li, T., Wei, G.~W., and Pan, F. (2021).
\newblock {Topological representations of crystalline compounds for the
  machine-learning prediction of materials properties}.
\newblock {\em npj Computational Materials}, 7(1):1--8.

\bibitem[Kaczynski et~al., 2006]{kaczynski2006}
Kaczynski, T., Mischaikow, K., and Mrozek, M. (2006).
\newblock {\em Computational homology}, volume 157.
\newblock Springer Science \& Business Media.

\bibitem[Kong et~al., 2013]{kong_blob}
Kong, H., Akakin, H.~C., and Sarma, S.~E. (2013).
\newblock {A generalized laplacian of gaussian filter for blob detection and
  its applications}.
\newblock {\em IEEE Transactions on Cybernetics}.

\bibitem[Kovalevsky, 1989]{kovalevsky1989}
Kovalevsky, V. (1989).
\newblock Finite topology as applied to image analysis.
\newblock {\em Computer Vision, Graphics, and Image Processing},
  46(2):141--161.

\bibitem[Lawrence et~al., 2021]{lawrence2021}
Lawrence, E.~L., Levin, B.~D., Boland, T., Chang, S.~L., and Crozier, P.~A.
  (2021).
\newblock Atomic scale characterization of fluxional cation behavior on
  nanoparticle surfaces: probing oxygen vacancy creation/annihilation at
  surface sites.
\newblock {\em ACS nano}, 15(2):2624--2634.

\bibitem[Lawrence et~al., 2019]{lawrence2019}
Lawrence, E.~L., Levin, B.~D., Miller, B.~K., and Crozier, P.~A. (2019).
\newblock {Approaches to Exploring Spatio-Temporal Surface Dynamics in
  Nanoparticles with in Situ Transmission Electron Microscopy}.
\newblock {\em Microscopy and Microanalysis}, 2(2020):86--94.

\bibitem[Lawson et~al., 2021]{lawson2021}
Lawson, A., Hoffman, T., Chung, Y.-M., Keegan, K., and Day, S. (2021).
\newblock {A density-based approach to feature detection in persistence
  diagrams for firn data}.
\newblock {\em Foundations of Data Science}.

\bibitem[Levin, 2021]{levin2021direct}
Levin, B.~D. (2021).
\newblock Direct detectors and their applications in electron microscopy for
  materials science.
\newblock {\em Journal of Physics: Materials}, 4(4):042005.

\bibitem[Levin et~al., 2020]{levin2020}
Levin, B.~D., Lawrence, E.~L., and Crozier, P.~A. (2020).
\newblock Tracking the picoscale spatial motion of atomic columns during
  dynamic structural change.
\newblock {\em Ultramicroscopy}, 213.

\bibitem[Lindeberg, 1990]{lindeberg1990}
Lindeberg, T. (1990).
\newblock Scale-space for discrete signals.
\newblock {\em IEEE transactions on pattern analysis and machine intelligence},
  12(3):234--254.

\bibitem[Lindeberg, 1998]{lindeberg1998}
Lindeberg, T. (1998).
\newblock Feature detection with automatic scale selection.
\newblock {\em International journal of computer vision}, 30(2):79--116.

\bibitem[Manzorro et~al., 2022]{xu2022}
Manzorro, R., Xu, Y., Vincent, J.~L., Rivera, R., Matteson, D.~S., and Crozier,
  P.~A. (2022).
\newblock Exploring blob detection to determine atomic column positions and
  intensities in time-resolved {TEM} images with ultra-low signal-to-noise.
\newblock {\em Microscopy and Microanalysis}, pages 1--14.

\bibitem[Massart, 1990]{massart1990}
Massart, P. (1990).
\newblock The tight constant in the dvoretzky-kiefer-wolfowitz inequality.
\newblock {\em The annals of Probability}, pages 1269--1283.

\bibitem[Mischaikow and Nanda, 2013]{mischaikow2013}
Mischaikow, K. and Nanda, V. (2013).
\newblock Morse theory for filtrations and efficient computation of persistent
  homology.
\newblock {\em Discrete \& Computational Geometry}, 50(2):330--353.

\bibitem[Mohan et~al., 2022]{mohan2020}
Mohan, S., Manzorro, R., Vincent, J.~L., Tang, B., Sheth, D.~Y., Simoncelli,
  E.~P., Matteson, D.~S., Crozier, P.~A., and Fernandez-Granda, C. (2022).
\newblock Deep denoising for scientific discovery: A case study in electron
  microscopy.
\newblock {\em IEEE Transactions on Computational Imaging}, 8:585--597.

\bibitem[Motta et~al., 2018]{motta2018}
Motta, F.~C., Neville, R., Shipman, P.~D., Pearson, D.~A., and Bradley, R.~M.
  (2018).
\newblock {Measures of order for nearly hexagonal lattices}.
\newblock {\em Physica D: Nonlinear Phenomena}, 380-381:17--30.

\bibitem[Mukherjee et~al., 2020]{mpfit2020}
Mukherjee, D., Miao, L., Stone, G., and Alem, N. (2020).
\newblock {mpfit: a robust method for fitting atomic resolution images with
  multiple Gaussian peaks}.
\newblock {\em Advanced Structural and Chemical Imaging}, 6(1):1--12.

\bibitem[Myllym{\"{a}}ki et~al., 2017]{myllymaki2017}
Myllym{\"{a}}ki, M., Mrkvi{\v{c}}ka, T., Grabarnik, P., Seijo, H., and Hahn, U.
  (2017).
\newblock {Global envelope tests for spatial processes}.
\newblock {\em Journal of the Royal Statistical Society: Series B (Statistical
  Methodology)}, 79(2):381--404.

\bibitem[Nakamura et~al., 2015]{nakamura2015}
Nakamura, T., Hiraoka, Y., Hirata, A., Escolar, E.~G., and Nishiura, Y. (2015).
\newblock Persistent homology and many-body atomic structure for medium-range
  order in the glass.
\newblock {\em Nanotechnology}, 26(30):304001.

\bibitem[Nord et~al., 2017]{atomap}
Nord, M., Vullum, P.~E., MacLaren, I., Tybell, T., and Holmestad, R. (2017).
\newblock Atomap: a new software tool for the automated analysis of atomic
  resolution images using two-dimensional gaussian fitting.
\newblock {\em Advanced structural and chemical imaging}, 3(1):1--12.

\bibitem[Rieck et~al., 2020]{rieck2020}
Rieck, B., Yates, T., Bock, C., Borgwardt, K., Wolf, G., Turk-Browne, N., and
  Krishnaswamy, S. (2020).
\newblock {Uncovering the topology of time-varying fMRI data using cubical
  persistence}.
\newblock {\em Advances in Neural Information Processing Systems (NeurIPS)},
  33(NeurIPS):6900--6912.

\bibitem[Robins et~al., 2011]{robins2011}
Robins, V., Wood, P.~J., and Sheppard, A.~P. (2011).
\newblock {Theory and algorithms for constructing discrete Morse complexes from
  grayscale digital images}.
\newblock {\em IEEE Transactions on pattern analysis and machine intelligence},
  33(8):1646--1658.

\bibitem[Robinson and Turner, 2017]{robinson2017}
Robinson, A. and Turner, K. (2017).
\newblock Hypothesis testing for topological data analysis.
\newblock {\em Journal of Applied and Computational Topology}, 1(2):241--261.

\bibitem[Rucco et~al., 2016]{rucco2016}
Rucco, M., Castiglione, F., Merelli, E., and Pettini, M. (2016).
\newblock Characterisation of the idiotypic immune network through persistent
  entropy.
\newblock In {\em Proceedings of ECCS 2014}, pages 117--128. Springer.

\bibitem[Ruskin et~al., 2013]{ruskin2013}
Ruskin, R.~S., Yu, Z., and Grigorieff, N. (2013).
\newblock Quantitative characterization of electron detectors for transmission
  electron microscopy.
\newblock {\em Journal of structural biology}, 184(3):385--393.

\bibitem[Skraba and Turner, 2020]{skraba2021wasserstein}
Skraba, P. and Turner, K. (2020).
\newblock Wasserstein stability for persistence diagrams.

\bibitem[Taylor et~al., 2007]{taylor2007maxima}
Taylor, J.~E., Worsley, K., and Gosselin, F. (2007).
\newblock Maxima of discretely sampled random fields, with an application to
  'bubbles'.
\newblock {\em Biometrika}, 94(1):1--18.

\bibitem[Thomas et~al., 2022]{supp}
Thomas, A.~M., Crozer, P.~A., Xu, Y., and Matteson, D.~S. (2022).
\newblock Supplement to ``{F}eature detection and hypothesis testing for
  extremely noisy nanoparticle images using topological data analysis''.

\bibitem[Turkes et~al., 2021]{turkes2021}
Turkes, R., Nys, J., Verdonck, T., and Latre, S. (2021).
\newblock {Noise robustness of persistent homology on greyscale images, across
  filtrations and signatures}.
\newblock {\em PLoS ONE}, 16(9 September):1--26.

\bibitem[Vandaele et~al., 2020]{vandaele2020}
Vandaele, R., Nervo, G.~A., and Gevaert, O. (2020).
\newblock {Topological image modification for object detection and topological
  image processing of skin lesions}.
\newblock {\em Scientific Reports}, 10(1):1--15.

\bibitem[Vejdemo-Johansson and Mukherjee, 2022]{vejdemo2020}
Vejdemo-Johansson, M. and Mukherjee, S. (2022).
\newblock Multiple hypothesis testing with persistent homology.
\newblock {\em Foundations of Data Science}, 4(4):667--705.

\bibitem[Vincent and Crozier, 2021]{vincent2021}
Vincent, J.~L. and Crozier, P.~A. (2021).
\newblock Atomic level fluxional behavior and activity of ceo2-supported pt
  catalysts for co oxidation.
\newblock {\em Nature communications}, 12(1):1--13.

\bibitem[Wagner et~al., 2012]{wagner2012}
Wagner, H., Chen, C., and Vu{\c{c}}ini, E. (2012).
\newblock Efficient computation of persistent homology for cubical data.
\newblock In {\em Topological methods in data analysis and visualization II},
  pages 91--106. Springer.

\end{thebibliography}


\begin{thebibliography}{6}
\providecommand{\natexlab}[1]{#1}
\providecommand{\url}[1]{\texttt{#1}}
\expandafter\ifx\csname urlstyle\endcsname\relax
  \providecommand{\doi}[1]{doi: #1}\else
  \providecommand{\doi}{doi: \begingroup \urlstyle{rm}\Url}\fi

\bibitem[Benjamini and Yekutieli(2001)]{benj_yeku}
Yoav Benjamini and Daniel Yekutieli.
\newblock {The control of the false discovery rate in multiple testing under
  dependency}.
\newblock \emph{The Annals of Statistics}, 29\penalty0 (4):\penalty0 1165 --
  1188, 2001.
\newblock \doi{10.1214/aos/1013699998}.
\newblock URL \url{https://doi.org/10.1214/aos/1013699998}.

\bibitem[Chen and Edelsbrunner(2011)]{chen2011diffusion}
Chao Chen and Herbert Edelsbrunner.
\newblock Diffusion runs low on persistence fast.
\newblock In \emph{2011 International Conference on Computer Vision}, pages
  423--430. IEEE, 2011.

\bibitem[Cohen-Steiner et~al.(2010)Cohen-Steiner, Edelsbrunner, Harer, and
  Mileyko]{cohensteiner2010}
David Cohen-Steiner, Herbert Edelsbrunner, John Harer, and Yuriy Mileyko.
\newblock Lipschitz functions have $l_p$-stable persistence.
\newblock \emph{Foundations of computational mathematics}, 10\penalty0
  (2):\penalty0 127--139, 2010.

\bibitem[Cressie(1993)]{cressie1993}
Noel Cressie.
\newblock \emph{Statistics for spatial data}.
\newblock John Wiley \& Sons, 1993.

\bibitem[Hiraoka and Tsunoda(2018)]{hiraoka_cube2018}
Yasuaki Hiraoka and Kenkichi Tsunoda.
\newblock {Limit Theorems for Random Cubical Homology}.
\newblock \emph{Discrete {\&} Computational Geometry}, 60\penalty0
  (3):\penalty0 665--687, oct 2018.
\newblock ISSN 0179-5376.
\newblock \doi{10.1007/s00454-018-0007-z}.
\newblock URL \url{http://link.springer.com/10.1007/s00454-018-0007-z}.

\bibitem[Rieck et~al.(2020)Rieck, Yates, Bock, Borgwardt, Wolf, Turk-Browne,
  and Krishnaswamy]{rieck2020}
Bastian Rieck, Tristan Yates, Christian Bock, Karsten Borgwardt, Guy Wolf,
  Nicholas Turk-Browne, and Smita Krishnaswamy.
\newblock {Uncovering the topology of time-varying fMRI data using cubical
  persistence}.
\newblock \emph{Advances in Neural Information Processing Systems (NeurIPS)},
  33\penalty0 (NeurIPS):\penalty0 6900--6912, 2020.
\newblock ISSN 10495258.

\end{thebibliography}
\bibliographystyle{apalike}

\end{document}


\title{\bf Supplementary materials for ``Feature Detection and Hypothesis Testing for Extremely Noisy Nanoparticle Images using Topological Data Analysis''}
  \author{Andrew M. Thomas \\
    Center for Applied Mathematics, Cornell University, \\
    Peter A. Crozier \\
    School for Engineering of Matter, Transport and Energy, ASU, \\
    Yuchen Xu \\ 
    Department of Statistics and Data Science, Cornell University, \\
    David S. Matteson \\
    Department of Statistics and Data Science, Cornell University}
\date{}
\maketitle

\section{Details on the collection of the primary image series}

By the primary image series, we mean the sequence of 1124 images in Section 5 of the paper. The data is a subset of a large dataset recorded as part of an in situ electron microscopy investigation of the behavior or Pt nanoparticles during catalysis, specifically CO oxidation (Vincent and Crozier 2021). The in situ TEM experiments explored the behavior of 1-5 nm Pt nanoparticles supported on $\text{CeO}_2$ cubes in a variety of different atmospheres. The experiments were performed on an FEI Titan environmental transmission electron microscope operating at 297 keV and were mostly carried out with an electron fluence of 600 e-/Å2/s. High resolution images were recorded with a Gatan K3 direct electron detector operating in CDS counting mode. Movies were recorded at up to 75 frames per second. In many cases the Pt particles were observed to randomly undergo dynamic structural reconfiguration. For this particular image sequence employed here, the catalyst was exposed to a CO partial pressure of $3 \times 10^{-2}$ Torr at room temperature. The video was entitled ``\texttt{PtCeO2\_6.tif}''.

\section{Additional checks on the assumptions of the primary image series}

\begin{figure}[h]
\centering
\includegraphics[width=4in]{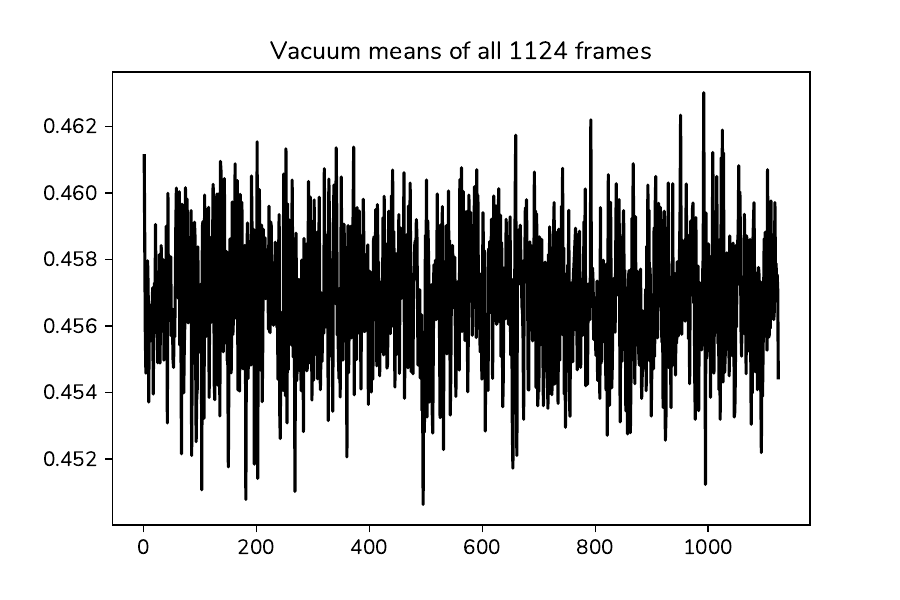}
\caption{Vacuum means for all 1124 frames.}
\label{f:vacmeans}
\end{figure}

The vacuum means for each frame in the primary image series of 1124 images are consonant with a stationary white noise series---see Figure~\ref{f:vacmeans}.

\begin{figure}[h]
\centering
\includegraphics[width=3in]{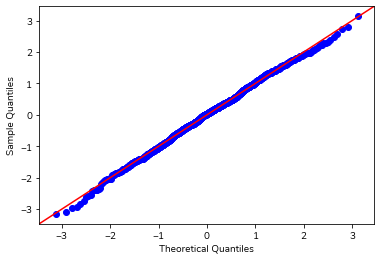}
\caption{Q-Q Plot of vacuum means for all 1124 frames.}
\label{f:qq_mean}
\end{figure}

Further reinforcing this notion is that the means of these frames are essentially normally distributed (after standardizing), as displayed in the q-q plot in Figure~\ref{f:qq_mean}. This follows if independence holds \emph{across} frames. 

\begin{figure}
\centering
\includegraphics[width=2.5in]{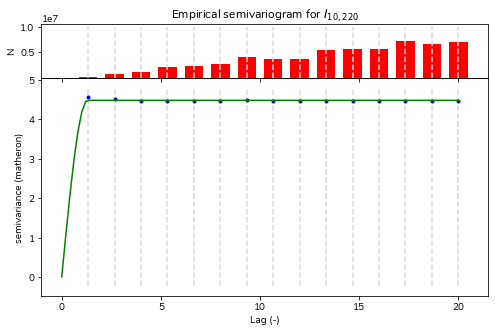}
\includegraphics[width=2.5in]{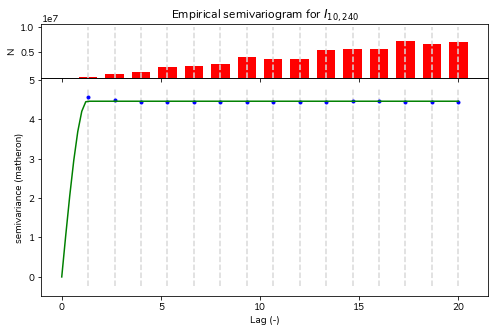}
\caption{Plots of empirical semivariograms for the vacuum regions of two images tested above, along with plotted estimated spherical model---see \cite{cressie1993}. The histograms are equal as the binning is the same and the collection of pairwise distances for the two $250\times400$ vacuum regions $U$ are identical.}
\label{f:semivariogram}
\end{figure}

An analysis of the semivariograms of the vacuum regions (Figure~\ref{f:semivariogram}) of the two images $I_{10, 220}$ and $I_{10, 240}$ indicates no violation of independence of the pixels \emph{within} frames. 

\section{Further results of the experiments and nonparametric setup}

For the $p$-values seen in the tables we used an earlier image series, consisting of the first 400 frames of the primary image series, and modulo some minor transformations---e.g. rotation, translation, etc. This video was entitled ``\texttt{010402\_stk\_400fr.tif}''. We used $F^{10}_U$ for the empirical cdf of pixel values. We see that there is little difference in Tables~\ref{t:mcpv} and \ref{t:mcpv2} from their parametric counterparts in the main document, especially with regards to the persistent entropy and ALPS statistic. 

For completeness, we conduct our Monte Carlo hypothesis test with a few different real-valued summaries of persistence diagrams, i.e. functionals $f\big(A(I_{m,\ell})\big)$, than ones in the main document. The first is signal-to-noise ratio (SNR), a summary that appears often in materials science, and defined as 
\[
\mathrm{SNR}\big(A(I)\big) := E(A(I))/\sqrt{M_2(A(I))},
\]
where more generally we have
\[
M_k(A(I)) = \frac{1}{|A(I)|} \sum_{(p,l)\in A(I)} (l - E(A(I)))^k.
\]
With this in mind, we can define the sample skewness as 
\[
S\big(A(I)\big) := \frac{M_3(A(I))}{M_2(A(I))^{3/2}}.
\]
It does not appear that either of these functional summaries of persistence diagrams appear in the literature. It seems that perhaps the negative of SNR and the skewness both perform well for small levels of smoothing, but the association breaks down as $\sigma$ goes from 2 to 4---see Tables~S3--S6. 

We also considered two more standard summaries, the lifetime sum (degree-1) and degree-2 total persistence \citep{cohensteiner2010}. These are generally defined (for $k \geq 1$) as
\[
L_k\big(A(I)\big) := \sum_{(p,l) \in A(I)} l^k  .
\]
Both the lifetime sum and the longest barcode were used in \cite{rieck2020}, with the longest barcode performing better in their prediction task. 

The functionals $L_k$ have desirable properties, such as stability in the sense that small perturbations of the underlying filtration yield small perturbations of $L_k$--see \cite{cohensteiner2010}. However, the $k$-norms $L_k(A(I))^{1/k}$ decay at the rate $O(1/\sigma)$ for $k \gtrapprox 5$ \citep{chen2011diffusion} which may explain the lack of utility (or consistency) of $L^1$ and $L^2$ across a wide range of $\sigma$. Furthermore, in the case of a cubical complex, the lifetime sum obeys a central limit theorem (cf. Theorem 2.13 in \citealt{hiraoka_cube2018}). Hence, future studies could investigate multiple testing using the lifetime sum in an ANOVA type setting. 

With that being said, the associations are quite sensitive to the smoothing parameter $\sigma$, just like the SNR and sample skewness. Hence, they were not considered in the paper. The Monte Carlo $p$-values using these summaries can be seen in Tables S3--S6. 

\begin{table}
\centering
\bgroup
\def\arraystretch{1.5}
\begin{tabular}{|l|r|r|r|r|r|}
\hline
Test statistic & $|A_{t(\sigma)}(I_{10, 240})|$ & $H\big(A(I_{10, 240})\big)$ & $L\big(A(I_{10, 240})\big)$ & $E\big(A(I_{10, 240})\big)$ & $\Delta\big(A(I_{10, 240})\big)$  \\
\hline
$\sigma = 2$ & 0.0001 & 0.0001 & 0.0587 & 0.1993 & 0.0007 \\
\hline
$\sigma = 4$ & 0.0001 & 0.0328 & 0.0056 & 0.0017 & 0.0001\\
\hline
$\sigma = 6$ & 0.1017 & 0.8109 & 0.5365 & 0.2476 & 0.093 \\
\hline
\end{tabular}
\egroup
\vspace{12pt}
\caption{Monte-Carlo $p$-values for various real-valued summaries $I_{10, 240}, \hat{I}_1, \dots, \hat{I}_{9999}$, with images generated $F^{10}_U$. That is, generated according to the product empirical pixel distribution.}
\label{t:mcpv} 
\vspace{12pt}

\bgroup
\def\arraystretch{1.5}
\begin{tabular}{|l|r|r|r|r|r|}
\hline
Test statistic & $|A_{t(\sigma)}(I_{10, 220})|$ & $H\big(A(I_{10, 220})\big)$ & $L\big(A(I_{10, 220})\big)$ & $E\big(A(I_{10, 220})\big)$ & $\Delta\big(A(I_{10, 240})\big)$ \\
\hline
$\sigma = 2$ & 0.0782 & 0.0572 & 0.1372 & 0.3118 & 0.0705 \\
\hline
$\sigma = 4$ & 0.1890 & 0.0230 & 0.0329 & 0.0027 & 0.0509 \\
\hline
$\sigma = 6$ & 0.2825 & 0.2104 & 0.1512 & 0.0294 & 0.1681 \\
\hline
\end{tabular}
\egroup
\vspace{12pt}
\caption{Monte-Carlo $p$-values for various real-valued summaries $I_{10, 220}, \hat{I}_1, \dots, \hat{I}_{9999}$, with images generated according to $F^{10}_U$. That is, generated according to the product empirical pixel distribution.}
\label{t:mcpv2}
\end{table}

\begin{table}
\centering
\bgroup
\def\arraystretch{1.5}
\begin{tabular}{|l|r|r|r|r|}
\hline
Test statistic & $\mathrm{SNR}(I_{10, 240})$ & $S\big(A(I_{10, 240})\big)$ & $L_1\big(A(I_{10, 240})\big)$ & $L_2\big(A(I_{10, 240})\big)$ \\
\hline
$\sigma = 2$ & 0.9802 & 0.0241 & 0.9999 & 0.2071 \\
\hline
$\sigma = 4$ & 0.8655 & 0.3939 & 0.0038 & 0.0025 \\
\hline
$\sigma = 6$ & 0.3250 & 0.8391 & 0.0838 & 0.1869 \\
\hline
\end{tabular}
\egroup
\vspace{12pt}
\caption{Monte Carlo $p$-values $p_n$ for various real-valued topological summaries of $I_{10, 240}, \hat{I}_1, \dots, \hat{I}_{9999}$ smoothed with varying values of $\sigma$. Images were generated according to $\hat{P}_{\ell, 0}$. With at least 95\% confidence the true $p$-value lies in $p_n \pm 0.0137$ (truncating at 0 or 1). These summaries where not considered in the main document. }
\label{t:poip_bad1} 
\vspace{12pt}

\bgroup
\def\arraystretch{1.5}
\begin{tabular}{|l|r|r|r|r|}
\hline
Test statistic & $\mathrm{SNR}(I_{10, 220})$ & $S\big(A(I_{10, 220})\big)$ & $L_1\big(A(I_{10, 220})\big)$ & $L_2\big(A(I_{10, 220})\big)$ \\
\hline
$\sigma = 2$ & 0.7440 & 0.0357 & 0.9006 & 0.5246 \\
\hline
$\sigma = 4$ & 0.1681 & 0.4042 & 0.2817 & 0.0523 \\
\hline
$\sigma = 6$ & 0.1532 & 0.6417 & 0.2733 & 0.157\\
\hline
\end{tabular}
\egroup
\vspace{12pt}
\caption{Monte Carlo $p$-values $p_n$ for various real-valued topological summaries of $I_{10, 220}, \hat{I}_1, \dots, \hat{I}_{9999}$ smoothed with varying values of $\sigma$. Images were generated according to $\hat{P}_{\ell, 0}$. With at least 95\% confidence the true $p$-value lies in $p_n \pm 0.0137$ (truncating at 0 or 1). These summaries where not considered in the main document.}
\label{t:poip_bad2}
\end{table}

\begin{table}[!ht]
\centering
\bgroup
\def\arraystretch{1.5}
\begin{tabular}{|l|r|r|r|r|}
\hline
Test statistic & $\mathrm{SNR}(I_{10, 240})$ & $S\big(A(I_{10, 240})\big)$ & $L_1\big(A(I_{10, 240})\big)$ & $L_2\big(A(I_{10, 240})\big)$   \\
\hline
$\sigma = 2$ & 0.9812 & 0.0231 & 0.9995 & 0.1573 \\
\hline
$\sigma = 4$ & 0.8701 & 0.3795 & 0.0026 & 0.0013 \\
\hline
$\sigma = 6$ & 0.3311 & 0.8375 & 0.0713 & 0.1784 \\
\hline
\end{tabular}
\egroup
\vspace{12pt}
\caption{Monte-Carlo $p$-values for various real-valued summaries $I_{10, 240}, \hat{I}_1, \dots, \hat{I}_{9999}$, with images generated according to $F^{10}_U$. That is, generated according to the product empirical pixel distribution. These summaries where not considered in the main document.}
\label{t:mcpv_bad1} 
\vspace{12pt}

\bgroup
\def\arraystretch{1.5}
\begin{tabular}{|l|r|r|r|r|}
\hline
Test statistic & $\mathrm{SNR}(I_{10, 220})$ & $S\big(A(I_{10, 220})\big)$ & $L_1\big(A(I_{10, 220})\big)$ & $L_2\big(A(I_{10, 220})\big)$ \\
\hline
$\sigma = 2$ & 0.7500 & 0.0368 & 0.8758 & 0.4552 \\
\hline
$\sigma = 4$ & 0.1692 & 0.4069 & 0.2527 & 0.0402 \\
\hline
$\sigma = 6$ & 0.1566 & 0.6462 & 0.2564 & 0.1433 \\
\hline
\end{tabular}
\egroup
\vspace{12pt}
\caption{Monte-Carlo $p$-values for various real-valued summaries $I_{10, 220}, \hat{I}_1, \dots, \hat{I}_{9999}$, with images generated according to $F^{10}_U$. That is, generated according to the product empirical pixel distribution. These summaries where not considered in the main document.}
\label{t:mcpv_bad2}
\end{table}

\begin{figure}
\centering
\includegraphics[height=1.5in]{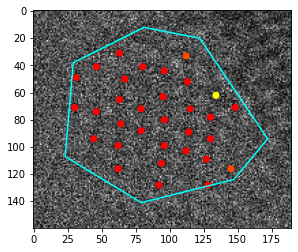}
\includegraphics[height=1.5in]{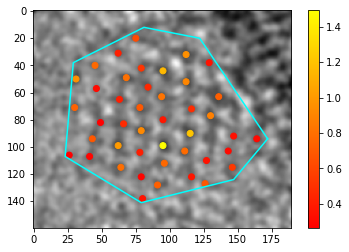}
\caption{Image $I_{10, 280}$ overlaid with output of Laplacian of Gaussian (LoG) blob detection centers (left) and same output of the cubical homology algorithm overlaid over smoothed $\sigma=2$ version of $I_{10,220}$. For such a high noise regime (see left image), LoG blob detection and our cubical persistence algorithm convey much of the same information. The algorithms (using \texttt{blob\_log} in scikit-image) run in roughly the same time, though the cubical homology algorithm is marginally faster. Colors in left image correspond to the bandwidth of the convolved Gaussian kernel.} 
\label{f:blob_compare}
\end{figure}

\section{Proofs of results}

We offer proofs of the results seen in the main document. First is a proof of an alternate representation of the ALPS statistic. 

\begin{proof}[Proof of Proposition 5.1]
Let $l_{(0)} \equiv 0$. By definition of Lebesgue integral and Tonelli's theorem, we have 
\begin{align*}
\int_0^{\infty} \log U(\eta) \dif{\eta} &= \int_0^{\infty} \log \bigg( \sum_{(x,l) \in A(I)} \ind{l > \eta} \bigg) \dif{\eta} \\
&= \int_{0}^{\infty} \sum_{i=0}^{n-1} \log (K-i) \ind{l_{(i)} < \eta \leq l_{(i+1)}} \dif{\eta} \\
&= \sum_{i=0}^{n-1} \big(l_{(i+1)} - l_{(i)}\big) \log (K-i) \\
&= \log \bigg(\prod_{i=0}^{K-1} (K-i)^{l_{(i+1)}} \bigg) - \log \bigg( \prod_{i=0}^{K-1} (K-i)^{l_{(i)}} \bigg).
\end{align*}
As $l_{(0)} = 0$, the product within the logarithm of the second term equals 
\begin{align*}
\prod_{i=1}^{K-1} \frac{(K-i)^{l_{(i)}}(K-i+1)^{l_{(i)}}}{(K-i+1)^{l_{(i)}}} &= \prod_{i=1}^{K-1} \Bigg(1 - \frac{1}{K-i+1} \Bigg)^{l_{(i)}} \prod_{i=1}^{K} (K-i+1)^{l_{(i)}} \\
&= \prod_{i=1}^{K-1} \Bigg(1 - \frac{1}{K-i+1} \Bigg)^{l_{(i)}} \prod_{i=0}^{K-1} (K-i)^{l_{(i+1)}},
\end{align*}
which finishes the proof.
\end{proof}

\begin{proof}[Proof of Proposition A.1]
We let $F_N$ be the empirical cdf of $t_1, \dots, t_N$---hence we may restate the test corresponding to Proposition A.1 as  
\[
p_{n,k} \leq \frac{F_N(t_k) \alpha}{C_N}.
\]
If $F_n$ is the empirical cdf of the $T_i$, it is straightforward to show that
\begin{equation} \label{e:gliv_cant}
\sup_{t \in \R} \big | F_n(t) - F(t) | \overset{a.s.}{\to} 0, \quad n \to \infty,
\end{equation}
implies that 
\begin{equation} \label{e:mcp_conv}
\sup_{t \in \R} \Bigg| \P(T \geq t) -  \frac{1}{n+1}\bigg( 1 + \sum_{i=1}^n \ind{T_i \geq t}\bigg) \Bigg| \overset{a.s.}{\to} 0 , \quad n \to \infty.
\end{equation}
The equation \eqref{e:gliv_cant} holds because of the Glivenko-Cantelli theorem. Suppose we denote 
\[
p_k := \P(T_i \geq t_k).
\]
Then \eqref{e:mcp_conv} implies that 
\[
\max_k |p_k - p_{n,k}| \overset{a.s.}{\to} 0, 
\]
as $n \to \infty$. Additionally, 
\[
\max_k \frac{|p_k - p_{n,k}|}{F_N(t_k)} \leq N \max_k |p_k - p_{n,k}| \overset{a.s.}{\to} 0, \quad n \to \infty.
\]
As $\P(\cup_k \{p_k/F_N(t_k) = \alpha/C_N\}) = 0$, we have that 
\[
\P\Big(\cap_k \{ \lim_{n \to \infty} \ind{p_{n,k}/F_N(t_k) \leq \alpha/C_N} = \ind{p_k/F_N(t_k) \leq \alpha/C_N } \Big) = 1.
\]
This implies that for $\P$-almost every $\omega \in \Omega$ that there exists some $M = M(\omega)$ such that if $n \geq M$ then 
\[ 
 \ind{p_{n,k}/F_N(t_k) \leq \alpha/C_N} =  \ind{p_k/F_N(t_k) \leq \alpha/C_N }, \text{ for all } k = 1, \dots, N.
\]
Let FDP be the false discovery proportion\footnote{The ratio of falsely rejected nulls over the total number of rejected nulls. It is defined to be 0 if no null hypotheses are rejected.} based on the test using $p_k$, $k = 1, \dots, N$ and let $\text{FDP}_n$ be the corresponding quantity for $p_{n,k}$. Note that the false discovery rate satisfies FDR = $\E[\text{FDP}]$ (and $\text{FDR}_n = \E[\text{FDP}_n]$). Based on the definition of $\text{FDP}$, for such an $\omega$, $\text{FDP}_n = \text{FDP}$ as the tests are identical. Hence,
\[
\text{FDP}_n \overset{a.s.}{\to} \text{FDP}, \quad n \to \infty.  
\]
If we let $\mathcal{T}_n$ be the natural filtration with respect to $T_1, \dots, T_n$ we have by the dominated convergence theorem for conditional expectations
\[
\text{FDR}_n = \E\big[\text{FDP}_n \mid \mathcal{T}_n\big] \overset{a.s.}{\to} \E\big[\text{FDP} \mid \mathcal{T}_n\big] = \text{FDR}. 
\]
Therefore based on the results of \cite{benj_yeku} for the false discovery rate $\text{FDR}$, we have
\[
\limsup_{n \to \infty} \text{FDR}_n \leq \alpha, \quad \text{a.s.} 
\]

\end{proof}

We now derive the confidence intervals for the $p$-values in the Tables S3 and S4 above and seen in Tables 2 and 3 in the main document. 

\begin{proof}
Let $t$ be our observed quantity of a test statistic $T$, and $T_1, \dots, T_n$ i.i.d. simulated values of the test statistic $T$ under then null hypothesis. Then if
\[
p_n(t) = \frac{1}{n+1}\bigg(1+\sum_{i=1}^n \ind{T_i \geq t}\bigg), 
\]
$p_n(t)$ is consistent estimator of the $p$-value $p(t) := P(T \geq t)$. In the case of a continuous $T$, we have that the empirical survival function
\[
\bar{F}_n(t) := \frac{1}{n} \ind{T_i > t},
\]
is also an unbiased estimator of $p(t)$, and we can use the DKW inequality to see that if 
\[
q_\alpha = \sqrt{-\frac{1}{2n}\log(\alpha/2)},
\]
then
\begin{equation}\label{e:qa}
\P\big( \sup_{t\in \R} |\bar{F}_n(t) - \bar{F}(t)| \leq q_\alpha \big) \geq 1-\alpha.
\end{equation}
Therefore, with \emph{at least} confidence level $1-\alpha$, we have that 
\[
\bar{F}_n(t) - q_\alpha \leq p(t) \leq \bar{F}_n(t) + q_\alpha,
\]
or 
\[
\bar{F}_n(t) \pm \sqrt{-\frac{1}{2n}\log(\alpha/2)}.
\]
Now, if $T$ is discrete, say on the integers, fix a $\delta \in (0,1]$ and suppose that $t$ is an integer. Then $\bar{F}_n(t-\delta)$ is a consistent estimator of $p(t)$. Thus, for $T$ discrete on the integers, we have
\[
\bar{F}_n(t-\delta) \pm \sqrt{-\frac{1}{2n}\log(\alpha/2)},
\]
is an at least $(1-\alpha)$\% confidence interval for $p(t)$. It is not difficult to show that
\[
\bar{F}_n(t) \leq p_n(t) + 1/n
\]
and likewise $\bar{F}_n(t-\delta) \leq p_n(t) + 1/n$ in the discrete case. Similarly, we can establish a lower bound for those terms of $p_n(t) - 1/n$. Hence, with at least $(1-\alpha)$\% confidence $p(t)$ lies within the interval 
\[
\Big[\max\{0, p_n(t) - q_{\alpha,n}\big\}, \min\{1, p_n(t) + q_{\alpha,n}\big\}\Big],
\]
where $q_{\alpha,n} := q_\alpha + 1/n$. The same interval estimate holds in the case that $T_i$ are generated according to some distribution with using a sufficient statistic for a parameter $\theta$ instead of $\theta$ due to invariance of \eqref{e:qa} when conditioning on a sufficient statistic for $\theta$.
\end{proof}

\section{Frames of interest in multiple testing demonstration}

\begin{figure}
    \centering
    \begin{subfigure}[b]{0.4\textwidth}
        \centering
        \includegraphics[width=2in]{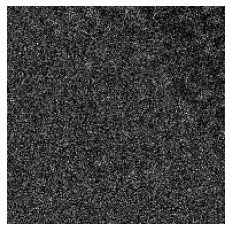}
        \caption{$I_{5, 176}$ ($k=35$), top region. \newline Lowest negative entropy frame. }
    \end{subfigure}
    \hfill
    \begin{subfigure}[b]{0.4\textwidth}
        \centering
        \includegraphics[width=2in]{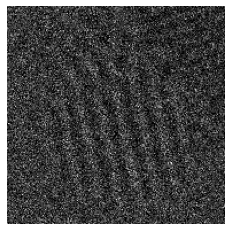}
        \caption{$I_{5, 1006}$ ($k=201$), top region. \newline Highest ALPS statistic frame. }
    \end{subfigure}
    \begin{subfigure}[t]{0.4\textwidth}
        \centering
        \includegraphics[width=2in]{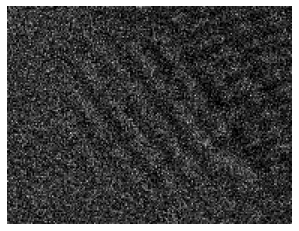}
        \caption{$I_{5, 101}$ ($k=20$), bottom region. \newline Highest negative entropy frame.}
    \end{subfigure}
    \hfill
    \begin{subfigure}[t]{0.4\textwidth}
        \centering
        \includegraphics[width=2in]{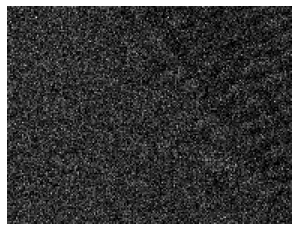}
        \caption{$I_{5, 861}$ ($k=172$), bottom region. \newline Lowest ALPS statistic frame.}
    \end{subfigure}
    \caption{Examples of frames in the image series $I_{5, 5k+1}$ that illustrate how the nanoparticles look when they take on low and high values of the ALPS statistic and persistent entropy.}
    \label{f:foi}
\end{figure}

Here we show a few frames corresponding to Figures 13 and 14 in the main document. We display the lowest negative entropy and highest ALPS statistic frames with $k=35$ and $k=201$, respectively for the top region and the highest negative entropy and lowest ALPS statistic frames for the bottom region with $k=20$ and $k=172$, respectively. These figures can be seen in Figure~\ref{f:foi}. 

\begin{figure}
    \centering
    \includegraphics[width=3in]{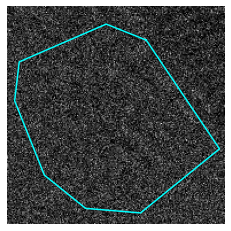}
    \caption{Barely significant at $\alpha=0.05$ level with respect to persistent entropy and non-significant with respect to the ALPS statistic, one can make out some faint atomic features in $I_{5, 761}$ for the top region. Boundaries of polygonal region $R$ have been superimposed on the image. }
    \label{f:i5761_top}
\end{figure}

Note that a frame is considered significant by our Benjamini-Hochberg procedure if in the top region the persistent entropy is $\leq 4.7537$ or the ALPs statistic is $\geq 2.3747$. The same values corresponding to the bottom region are $\leq 3.7868$ and $\geq 1.8943$. We now examine two frames which are significant with respect to persistent entropy but not with respect to the ALPS statistic. The first is $I_{5, 761}$ ($k=152$) for the top region, where $H\big(A(I_{5, 761})\big) = 4.6924 \leq 4.7537$ and $\Delta\big(A(I_{5, 761})\big) = 2.1725 \not \geq 2.3747$, which can be seen in Figure~\ref{f:i5761_top}. 

\begin{figure}
    \centering
    \includegraphics[width=3in]{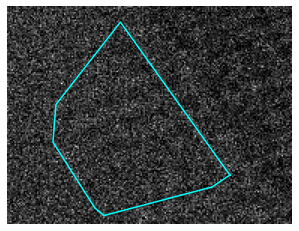}
    \caption{Barely significant at $\alpha=0.05$ level with respect to persistent entropy and  barely non-significant with respect to the ALPS statistic, one can make out some faint atomic features in $I_{5, 501}$ for the bottom region. Boundaries of polygonal region $R$ have been superimposed on the image. }
    \label{f:i5501_bottom}
\end{figure}

The second is the image $I_{5, 501}$ ($k=100$) for the bottom nanoparticle region; it is an image which shares significance with respect to persistent entropy $H\big(A(I_{5, 761})\big) = 3.7654 \leq 3.7868$ and non-significance with respect to the ALPS statistic. However, it is quite closer in absolute terms to the ALPS statistic threshold $\Delta\big(A(I_{5, 501})\big) = 1.8728 \not \geq 1.8943$. The intuitive interpretation is perhaps the ALPS statistic is closer to capturing the ``intuitive'' nature of significance better than persistent entropy. Indeed, there seems to be more signal in Figure~\ref{f:i5501_bottom} than in Figure~\ref{f:i5761_top}.

\bibliography{AtomSurfTDA}
\bibliographystyle{plainnat}